\documentclass[pra,twocolumn,groupedaddress,superscriptaddress,amsmath,amssymb,a4paper]{revtex4}
\usepackage{geometry}
\usepackage{graphicx}
\usepackage{mathrsfs}
\usepackage{amssymb}
\usepackage{amsmath}
\usepackage{amsthm}
\usepackage{amsbsy}
\usepackage{bm}
\usepackage{hyperref}
\usepackage[T1]{fontenc}

\theoremstyle{definition}

\newtheorem{example}{Example}

\newcommand{\bra}[1]{\langle #1|}
\newcommand{\ket}[1]{| #1 \rangle }

\newcommand{\blockmatrix}[3]{
\begin{minipage}[t][#2][c]{#1}%
\center%
#3%
\end{minipage}%
}%

\begin{document}

\title{Divisibility of quantum dynamical maps and collision models}

\author{S. N. Filippov}

\affiliation{Moscow Institute of Physics and Technology,
Institutskii Per. 9, Dolgoprudny, Moscow Region 141700, Russia}

\affiliation{Institute of Physics and Technology, Russian Academy
of Sciences, Nakhimovskii Pr. 34, Moscow 117218, Russia}

\author{J. Piilo}

\affiliation{Turku Centre for Quantum Physics, Department of
Physics and Astronomy, University of Turku, FI-20014, Turun
Yliopisto, Finland}

\author{S. Maniscalco}

\affiliation{Turku Centre for Quantum Physics, Department of
Physics and Astronomy, University of Turku, FI-20014, Turun
Yliopisto, Finland}

\affiliation{Centre for Quantum Engineering, Department of Applied Physics, School of Science, Aalto University, P.O. Box 11000, FI-00076 Aalto, Finland}

\author{M. Ziman}

\affiliation{Institute of Physics, Slovak Academy of Sciences,
D\'{u}bravsk\'{a} cesta 9, Bratislava 84511, Slovakia}

\affiliation{Faculty of Informatics, Masaryk University,
Botanick\'{a} 68a, Brno 60200, Czech Republic}

\begin{abstract}
Divisibility of dynamical maps is visualized by trajectories in
the parameter space and analyzed within the framework of collision
models. We introduce ultimate completely positive (CP) divisible
processes, which lose CP divisibility under infinitesimal
perturbations, and characterize Pauli dynamical semigroups
exhibiting such a property. We construct collision models with
factorized environment particles, which realize additivity and
multiplicativity of generators of CP divisible maps. A mixture of
dynamical maps is obtained with the help of correlated
environment. Mixture of ultimate CP divisible processes is shown
to result in a new class of eternal CP indivisible evolutions. We
explicitly find collision models leading to weakly and essentially
non-Markovian Pauli dynamical maps.
\end{abstract}

\maketitle

\section{Introduction}

Theory of open quantum systems studies dynamical maps $\Phi_t$
that naturally occur when the system in question interacts with its
environment. Dynamical maps are the key objects in the analysis of
quantum information transmission through noisy
channels~\cite{holevo-giovannetti-2012} and quantum information
processing in real systems~\cite{palma-1996}. The effect of open
system dynamics on quantum entanglement and entanglement-based
information protocols is reviewed, e.g., in
Ref.~\cite{aolita-2015}. Over the last decade quantum dynamical
maps were intensively studied with respect to characterization of
their non-Markovian
behavior~\cite{rivas-2014,breuer-2016,de-vega-2017} and its
experimental
observation~\cite{liu-2011,tang-2012, chiuri-2012,groblacher-2015,bernardes-2015,bernardes-2016, cialdi-2017}.
Quantitative approaches to non-Markovianity include non-monotonic
distinguishability of states~\cite{breuer-2009,laine-2010},
different divisibilities of dynamical
maps~\cite{wolf-prl-2008,rivas-2010}, monitoring the volume of
accessible states~\cite{lorenzo-2013} and
others~\cite{lu-2010,luo-2012,bylicka-2014,dhar-2015}. In
particular, the divisibility approach is based on the
decomposition property $\Phi_{t+s} = \Theta_{t,t+s} \Phi_t$ and
explores features of the intermediate map
$\Theta_{t,t+s}$~\cite{wolf-2008}. Various types of divisibility
induce alternative measures to quantify non-Markovianity, however,
one should be careful with the physical interpretation of memory
effects~\cite{megier-2016}.

From mathematical viewpoint, the open system dynamics in the
Schr\"{o}dinger picture is given by the transformation $\varrho(t)
= \Phi_t [\varrho(0)]$, where $\Phi_t$ is the dynamical map
(process) that is a one-parameter family of completely positive
trace preserving (CPT) maps, $t \geqslant 0$ is the evolution
time, $\Phi_0 = {\rm Id}$, the identity transformation. The
dilation of the dynamical map is
\begin{equation}
\label{dilation} \Phi_t [\varrho] = {\rm tr}_{\rm env} \left\{ U_t
(\varrho \otimes \xi) U_t^{\dag} \right\},
\end{equation}

\noindent where $U_t$ is the unitary evolution of the system and
the environment, $\xi$ is the initial state of environment.

Complete positivity (CP) of $\Phi_t$ means that the map $\Phi_t
\otimes {\rm Id}_k$ is positive for all identity transformations
${\rm Id}_k$ of $k$-level ancillary systems, which can be
potentially entangled with the system in question. If
$\Theta_{t,t+s}$ is CP for all $t$ and $s \neq t$, then the
process $\Phi_t$ is called \emph{CP divisible}. Such a definition of CP
divisibility is a global-in-time property of the whole family
$\{\Phi_t\}_{t \geqslant 0}$. In contrast, to underline the
time-local behavior, we will refer to a process $\Phi_t$ as CP
divisible at time $t_0$ if there exists $s_0 > 0$ such that
$\Theta_{t_0,t_0+s}$ is CP for all $s \in (0, s_0)$. If the
dynamical map $\Phi_t$ is not CP divisible for all time moments $t
\geqslant 0$, then $\Phi_t$ is called \emph{eternal CP
indivisible}~\cite{hall-2014}. CP divisibility of a bijective
dynamical map was shown to be equivalent to the distinguishability
of states in the extended Hilbert space \cite{bylicka-2016}.

Replacing CP by any other property [viz. positivity (P),
$k$-positivity, volume of accessible states, etc.] we obtain
definitions of the global and time-local divisibility properties
of the dynamical map $\Phi_t$. Processes, which are not CP
divisible but are P divisible, are also called \emph{weakly
non-Markovian}, whereas P indivisible processes are called
\emph{essentially non-Markovian}~\cite{chruscinski-maniscalco-2014}.

Since any linear map $\Phi$ between finite dimensional spaces can
be defined by a set of real parameters $\boldsymbol{\lambda} =
\lambda_1,\ldots,\lambda_n$, any smooth process $\Phi_t$ is then
determined by a continuous trajectory $\boldsymbol{\lambda}(t)$ in
the parameter space. Such a trajectory provides a pictorial
representation of the dynamical map in $\mathbb{R}^n$, which is
particularly visual in the case of qubit Pauli maps given by 3
parameters (see,
e.g.,~\cite{hall-2008,hall-2014,nalezyty-2015,wudarski-2015,chruscinski-siudzinska-2016}).
Analyzing the process trajectory in the parameter space, one can
not only get an intuition about the quantum dynamics (for
instance, by observing the Bloch ball transformation for qubit
dynamics) but also reveal its divisibility properties. The first
goal of this paper is to describe different forms of Markovian and
non-Markovian Pauli dynamical maps in terms of trajectories in the
parameter space.

Pictorial representation of some dynamical map $\Phi_t$ in the
form of trajectory $\boldsymbol{\lambda}(t)$ raises a question of
stability of the process with respect to a continuous (infinitely
differentiable) trajectory perturbation $\boldsymbol{\lambda}(t)
\rightarrow \boldsymbol{\lambda}(t) + \delta
\boldsymbol{\lambda}(t)$, with the perturbed map $\Phi_t + \delta
\Phi_t$ being a valid quantum dynamical evolution. A process
$\Phi_t$, which is originally CP divisible at time $t_0$, may lose
the property of being CP divisible at this time due to a
time-local perturbation $\delta\boldsymbol{\lambda}(t)$ such that
$\delta\boldsymbol{\lambda}(t) = 0$ if $t \leqslant t_0$. If this
is the case, $\Phi_t$ is called \emph{ultimate} CP divisible at
time $t_0$. There exist processes $\Phi_t$ that are ultimate CP
divisible for all time moments $t \geqslant 0$. We fully
characterize Pauli dynamical semigroups exhibiting such a
property.

In this paper, we show that the mathematical concepts of
divisibility are closely related with the underlying physical
models of quantum dynamical maps. From physical viewpoint, any
dynamical map $\Phi_t$ can be seen as a simplified description of
the system-environment evolution with no regard to the environment
structure and particular microscopic interactions between
environment quanta and the system. Many dynamical maps can be
derived under some assumptions (weak coupling, low density, etc.)
from a microscopic system-environment Hamiltonian and particular
state of the
environment~\cite{breuer-petruccione-2002,de-vega-2017}. In our
analysis, we will resort to so-called collision models in which
the motional and internal degrees of freedom can be considered
separately: the motion Hamiltonian determines a sequence of
collisions with environment particles, and the system-environment
interaction Hamiltonian becomes significant during collisions and
affects internal degrees of freedom of the system and an impacted
environment particle. Relaxation mechanism via such a ``stirring''
process was first considered in Ref.~\cite{rau-1963}.
Thermalization, homogenization of the system to a particular
state, and pure dephasing were simulated via a collision model
with identical uncorrelated environment particles in
\cite{scarani-2002,ziman-2002,ziman-buzek-2011}. Even if
environment particles are uncorrelated originally, they become
partially correlated (entangled) with the system during
collisions, so such an environment exhibits memory effects for
further systems interacting with it
\cite{kretschmann-2005,giovannetti-2005,giovannetti-2012,caruso-2014}.
Moreover, environment particles may be initially correlated
(quantumly or classically) due to interactions between each other
as it takes place in solids and quantum gases, and such
correlations may result in non-Markovian
dynamics~\cite{rybar-2012,bernardes-2014,bernardes-2016,bernardes-2017,dabrowska-2017}.
Non-Markovian effects also appear in collision models, where the
system can interact with the same environment particle several
times~\cite{pellegrini-2009,bodor-2013}, or an environment
particle impacted by a system collides with another environment
particle, which later collides with the
system~\cite{ciccarello-2013,ciccarello-ps-2013,mccloskey-2014,kretschmer-2016}.
The latter scheme is equivalent to a scenario, when the quantum
system in question is coherently coupled to an auxiliary system
interacting with Markovian bath via
collisions~\cite{budini-2013,lorenzo-2016}. Collision models
adequately describe a particle in semi-quantal spin
gases~\cite{hartmann-2005,koniorczyk-2008}, a
micromaser~\cite{vacchini-2016}, a two-level system that interacts
with spatio-temporal modes passing through it only
once~\cite{diosi-2012}, and more complex systems with involved
interaction
graphs~\cite{lorenzo-2017,cusumano-2017,lorenzo-2-2017} as well as
experiments with an engineered environment in nuclear magnetic
resonance~\cite{bernardes-2016} and in photonic
systems~\cite{lupo-2010,bernardes-2015,jin-2015}. Collision models
were also exploited in the microscopic description of Landauer's
principle~\cite{lorenzo-2015}.

In the appropriate continuous limit of infinitesimal interaction
time $\tau \rightarrow 0$, the collision model describes a smooth
dynamical map $\Phi_t$~\cite{attal-2006,lorenzo-2017}. Even if we
consider simple interaction graphs, when the system interacts with
each environment particle only once, collision models successfully
simulate dynamical processes $\Phi_t$ with different
divisibility properties~\cite{rybar-2012}. So we resort to a collision model
with generally correlated environment particles, the correlations
being attributed to prior interactions among environment
constituents.

The second goal of this paper is to demonstrate that the
divisibility property of the dynamical map $\Phi_t$ is closely
related with the collision model describing it. Clearly, any CP
divisible dynamics can be obtained with uncorrelated (factorized)
environment states. We fully characterize ultimate CP divisible
Pauli dynamical semigroups and corresponding collision models. CP
indivisible dynamics necessarily involves correlations among
environment particles.

Surprisingly, a convex sum $p_1\Phi_t^{(1)} + p_2\Phi_t^{(2)} +
\ldots$ of CP divisible processes
$\Phi_t^{(1)},\Phi_t^{(1)},\ldots$ can exhibit eternal CP
indivisibility, for instance, this takes place for the convex sum
of two dephasing dynamical maps~\cite{hall-2014,megier-2016}. We
provide new families of eternal CP indivisible processes and
construct a collision model with correlated environment, which
simulates them.

In contrast to a convex sum of dynamical maps, a conical
(weighted) combination $\alpha \mathcal{L}_t^{(1)} +
\beta\mathcal{L}_t^{(2)}$ of time-dependent generators
$\mathcal{L}_t^{(1)}$ and $\mathcal{L}_t^{(2)}$ does not
necessarily represent a valid
generator~\cite{kolodynski-2017,benatti-2017} unless master
equations $\frac{\partial \varrho}{\partial t} =
\mathcal{L}_t^{(1)} [\varrho]$ and $\frac{\partial
\varrho}{\partial t} = \mathcal{L}_t^{(2)} [\varrho]$ both define
CP divisible processes. When the latter condition is
fulfilled, we demonstrate a collision model realizing the master
equation $\frac{\partial \varrho}{\partial t} = \alpha
\mathcal{L}_t^{(1)} [\varrho] + \beta\mathcal{L}_t^{(2)}
[\varrho]$ for arbitrary non-negative weight coefficients $\alpha$
and $\beta$.

The paper is organized as follows. In
Sec.~\ref{section-pictorial}, we review divisibility properties of
Pauli dynamical maps in pictorial representation. In
Sec.~\ref{section-ultimate}, ultimate CP divisible semigroups are
studied. In Sec.~\ref{section-collision-semigroup}, we provide a
general collision model for ultimate CP divisible Pauli processes.
In Sec.~\ref{section-multiplicativity}, we demonstrate collision
models that realize multiplicativity and additivity of time-local
generators for CP divisible processes. In
Sec.~\ref{section-mixtures}, we construct a correlated environment
which leads to a mixture of CP divisbible processes. In
Sec.~\ref{section-eternal}, a new two-parameter family of eternal
CP indivisible Pauli maps is presented. In
Secs.~\ref{section-p-divisibility}
and~\ref{section-p-indivisibility-all}, we review P divisible and
P indivisible processes, respectively, as well as the physics of
underlying collision models. In Sec.~\ref{section-arbitrary}, we
provide a constructive collision model for an arbitrary Pauli
dynamical map $\Phi_t$. In Sec.~\ref{section-volume}, we discuss
dynamical maps which continuously shrink the volume of accessible
states, however, are not P divisible. In
Sec.~\ref{section-conclusions}, brief conclusions are given.

\section{Divisibility of Pauli maps in pictorial representation}
\label{section-pictorial}

\begin{figure}
\includegraphics[width=8.5cm]{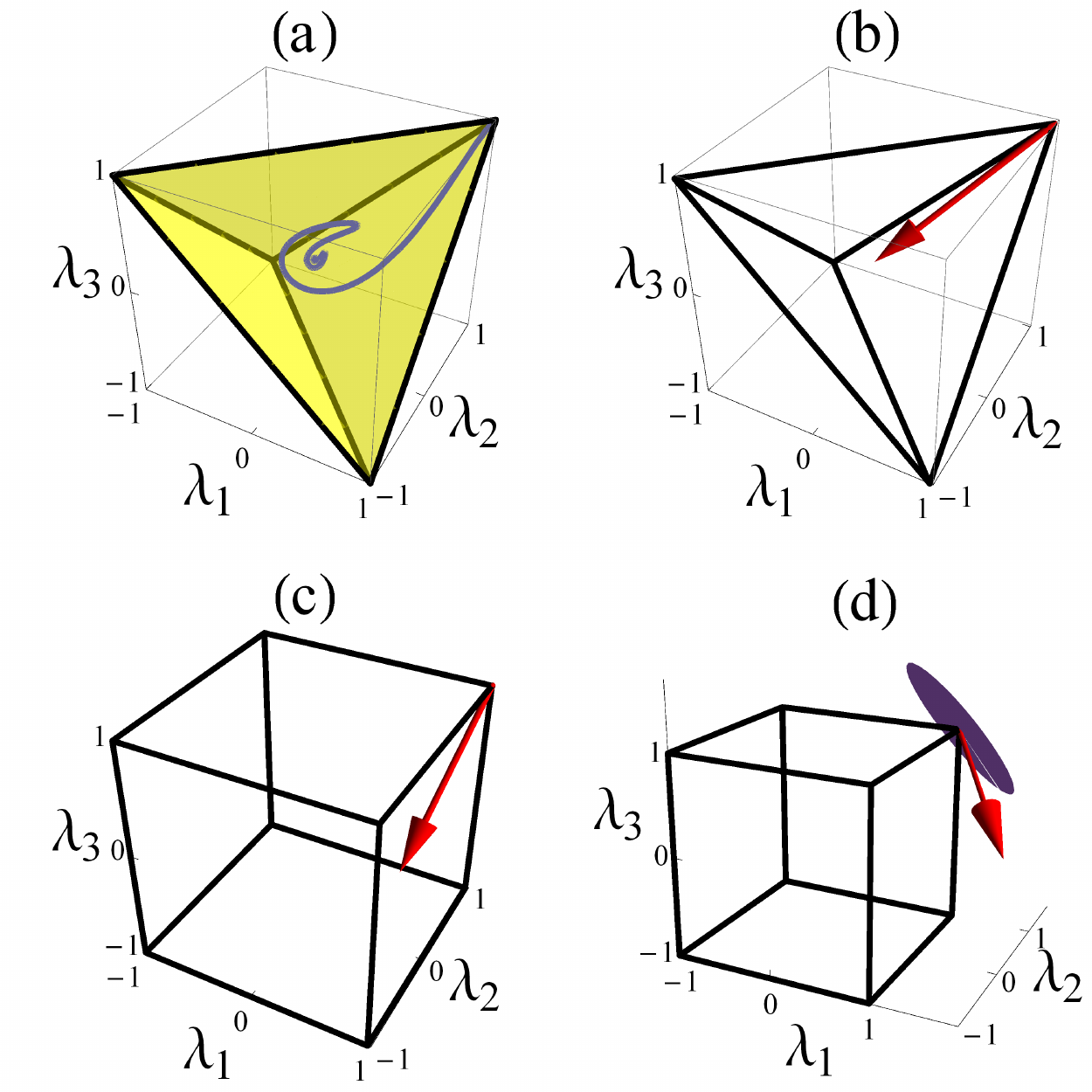}
\caption{\label{figure1} (a) Physical Pauli processes correspond
to curves inside the set of completely positive maps. The direction of
the vector $\boldsymbol{\kappa}$ given by Eq.~\eqref{kappa}
defines the properties of Pauli dynamical maps: (b) CP divisibility,
(c) P divisibility, (d) monotonic shrink of the volume of
accessible states.}
\end{figure}

A trajectory $\boldsymbol{\lambda}(t)$ becomes particularly visual
for Pauli qubit processes $\Phi_t:{\cal B}({\cal H}_2)\mapsto
{\cal B}({\cal H}_2)$ that are characterized by three real
parameters $\lambda_1(t),\lambda_2(t),\lambda_3(t)$ as follows:
\begin{equation}
\label{unital-process} \Phi_t[\varrho] = \frac{1}{2} \left( {\rm
tr}[\varrho] I + \sum_{j=1}^{3}\lambda_{j}(t) {\rm tr}[\sigma_{j}
\varrho] \sigma_{j} \right),
\end{equation}

\noindent where $\sigma_1,\sigma_2,\sigma_3$ is a conventional set
of Pauli operators. The map $\Phi_t$ is known to be positive if
$-1 \leqslant \lambda_1(t),\lambda_2(t),\lambda_3(t) \leqslant 1$
(cube in the parameter space) and completely positive if $1 \pm
\lambda_3(t) \geqslant |\lambda_1(t) \pm \lambda_2(t)|$
(tetrahedron in the parameter space)
\cite{ruskai-2002,bengtsson-2006}. In the case of a general
physical evolution $\Phi_t$ with initially factorized system and
environment, the trajectory $\boldsymbol{\lambda}(t)$ can be an
arbitrary smooth curve inside the tetrahedron $1 \pm \lambda_3(t)
\geqslant |\lambda_1(t) \pm \lambda_2(t)|$ (see
Fig.~\ref{figure1}a).

Suppose the map $\Phi_t$ is invertible and $s$ tends to zero, then
\begin{eqnarray}
\label{Phi-t-s} && \Phi_{t+s}[\varrho(0)] = \Phi_t[\varrho(0)] + s
\dot{\Phi}_t [\varrho(0)] \nonumber\\ && =  \varrho(t) + s
\dot{\Phi}_t \circ \Phi_t^{-1} [\varrho(t)] =  \Theta_{t,t+s}
[\varrho(t)],
\end{eqnarray}

\noindent where $\dot{\Phi}_t = \frac{\partial}{\partial t}
\Phi_t$. From Eq.~\eqref{Phi-t-s} it follows that
\begin{equation}
\Theta_{t,t+s} = {\rm Id} + s \dot{\Phi}_t \circ \Phi_t^{-1}
\end{equation}
as $s \rightarrow 0$.

The map $\dot{\Phi}_t \circ \Phi_t^{-1}$ defines a direction in
the parameter space in which the process progresses. Using the
explicit form of Eq.~\eqref{unital-process} we get
\begin{equation}
\dot{\Phi}_t \circ \Phi_t^{-1}[X] = \frac{1}{2} \sum_{j=1}^{3}
\frac{\dot{\lambda}_{j}(t)}{\lambda_{j}(t)} {\rm tr}[\sigma_{j} X]
\sigma_{j},
\end{equation}

\noindent which identifies the vector
\begin{equation}
\label{kappa} \boldsymbol{\kappa}(t) = \left(
\frac{\dot{\lambda}_1(t)}{\lambda_1(t)},
\frac{\dot{\lambda}_2(t)}{\lambda_2(t)},
\frac{\dot{\lambda}_3(t)}{\lambda_3(t)} \right)
\end{equation}
\noindent representing the dynamical map
in the parameter space of qubit unital channels.
Let us stress that $\boldsymbol{\kappa}(t)$ is not a tangent line
to the trajectory $\boldsymbol{\lambda}(t)$. Such vector
$\boldsymbol{\kappa}(t)$ can be drawn at any time moment
$t$ for a sufficiently smooth trajectory $\boldsymbol{\lambda}(t)$
in the parameter space, making the divisibility property more
apparent.

In particular, if $s \rightarrow 0$, then the map $\Theta_{t,t+s}$ is completely
positive if and only if the vector $\boldsymbol{\kappa}(t)$ drawn
from the corner $(1,1,1)$ of the parameter space points inside the
tetrahedron of completely positive maps in Fig.~\ref{figure1}b,
i.e. the scalar products of $\boldsymbol{\kappa}(t)$ with vectors
$(-1,1,1)$, $(1,-1,1)$, and $(1,1,-1)$ are all non-positive:
\begin{eqnarray}
&& \label{CP-1} - \kappa_1(t) + \kappa_2(t) + \kappa_3(t) \leqslant 0, \\
&& \label{CP-2} \kappa_1(t) - \kappa_2(t) + \kappa_3(t) \leqslant 0, \\
&& \label{CP-3} \kappa_1(t) + \kappa_2(t) - \kappa_3(t) \leqslant 0.
\end{eqnarray}

Analogously, if $s \rightarrow 0$, then the map $\Theta_{t,t+s}$
is positive if and only if the vector $\boldsymbol{\kappa}(t)$ drawn
from the corner $(1,1,1)$ of the parameter space points inside the
cube of positive maps in Fig.~\ref{figure1}c, i.e. the scalar
products of $\boldsymbol{\kappa}(t)$ with vectors $(1,0,0)$,
$(0,1,0)$, and $(0,0,1)$ are all non-positive:
\begin{equation}
  \kappa_1(t) \leqslant 0, \quad \kappa_2(t) \leqslant 0, \quad
  \kappa_3(t)\leqslant 0.
\end{equation}

For the uniform measure of qubit states inside the Bloch ball
(metric induced by Hilbert--Schmidt
distance~\cite{bengtsson-2006}) the volume of accessible states
for the Pauli map is
$V(t)=|\lambda_1(t)\lambda_2(t)\lambda_3(t)|$. The map
$\Theta_{t,t+s}$ shrinks the volume of accessible states if and
only if $\prod_{i=1}^3 (1 + s\kappa_i(t)) \leqslant 1$, which in the
limit $s \rightarrow 0$ transforms into requirement
\begin{equation}
\kappa_1(t) + \kappa_2(t) + \kappa_3(t) \leqslant 0.
\end{equation}

\noindent Here we have taken into account that if $\kappa_1 +
\kappa_2 + \kappa_3 = 0$ and at least one $\kappa_i \neq 0$, then
$\kappa_1\kappa_2 + \kappa_2\kappa_3 + \kappa_3\kappa_1 = -
\frac{1}{2} (\kappa_1^2 + \kappa_2^2 + \kappa_3^2) < 0$, which
implies $\prod_{i=1}^3 (1 + s\kappa_i) < 1$. Geometrically, the
vector $\boldsymbol{\kappa}$(t) has non-positive scalar product with
the vector $(1,1,1)$, i.e. the vector $\boldsymbol{\kappa}(t)$ drawn
from the corner $(1,1,1)$ in parameter space points to a specific
half-space separated by the plane $\lambda_1 + \lambda_2 +
\lambda_3 = 3$ (see Fig.~\ref{figure1}d).

\section{Ultimate CP divisibility of semigroup dynamics}
\label{section-ultimate}

Consider a semigroup dynamics $\Phi_t = e^{\mathcal{L} t}$, where
$\mathcal{L}: \mathcal{B}(\mathcal{H}_2) \mapsto
\mathcal{B}(\mathcal{H}_2)$ is a time-independent
generating map of the form~\cite{gks-1976,lindblad-1976}
\begin{equation}
\mathcal{L}[\varrho] = -i[H,\varrho] + \sum_{k} \gamma_k \left(
A_k \varrho A_k^{\dag} - \frac{1}{2} \{ \varrho, A_k^{\dag} A_k \}
\right),
\end{equation}

\noindent where $H$ is Hermitian and $\gamma_k \geqslant 0$. It
follows that for semigroup dynamics the identity
$\Phi_{t+s}=\Phi_t\circ\Phi_s$ holds for all $t,s\geq 0$.
Consequently $\Theta_{t,t+s} =\Phi_s = e^{\mathcal{L} s}$, hence,
the semigroup dynamics is always CP divisible.

The time evolution of the density operator is given by equation
\begin{equation}
\label{lindblad} \frac{\partial \varrho}{\partial t} = \mathcal{L}
[\varrho].
\end{equation}
Consider now an infinitesimal perturbations of
Eq.~\eqref{lindblad}
\begin{equation}
\label{lindblad-perturbed} \frac{\partial \varrho}{\partial t} =
\big( \mathcal{L} + \delta\mathcal{L}_t \big)[\varrho],
\end{equation}

\noindent where $\delta\mathcal{L}_0=0$. The term
$\delta\mathcal{L}_t$ describes an infinitely differentiable
deviation from dynamics \eqref{lindblad}, and can be attributed
to, e.g., a slightly modified environment or a fluctuating
interaction between system and environment. By definition
we say that a semigroup dynamics is \emph{ultimate CP divisible}
if it becomes CP indivisible under some perturbation
$\delta\mathcal{L}$.

For qubit unital semigroup processes \eqref{unital-process}
we have $\lambda_j(t) = e^{-\Gamma_j t}$ and, consequently,
the vector $\boldsymbol{\kappa}(t)=-\boldsymbol{\Gamma}$
is time-independent and we used
$\boldsymbol{\Gamma}=(\Gamma_1,\Gamma_2,\Gamma_3)$.
By definition the perturbations $\delta\mathcal{L}_t$
are introducing only minor changes and the deviated vector $\boldsymbol{\kappa}
+ \delta\boldsymbol{\kappa}$ would satisfy Eqs.~\eqref{CP-1}--\eqref{CP-3}
whenever these inequalities for the unperturbed case $\boldsymbol{\kappa}$
are strict. It turns out that qubit unital semigroup dynamics can be
ultimately CP only if $\kappa_i+\kappa_j-\kappa_k=0$ for some
permutation of indexes $i,j,k\in\{1,2,3\}$. In fact, in such case
there exists an infinitesimal perturbation $\delta\mathcal{L}_t$ resulting
in a dynamical map $\Phi_t + \delta\Phi_t$ violating
Eqs.~\eqref{CP-1}--\eqref{CP-3}. Taking into account the definition
of $\boldsymbol{\kappa}$ we find that the condition
$\kappa_i+\kappa_j-\kappa_k=0$ translates into differential
equation $\frac{d}{dt}  \ln (\lambda_i \lambda_j) =
\frac{d}{dt}  \ln (\lambda_k)$ with the solution
$\lambda_i(t) \lambda_j(t) = c\lambda_k(t)$, where the constant
$c$ can be found from the
initial condition $\lambda_1(0) = \lambda_2(0) = \lambda_3(0) =1$.
In conclusion, $c=1$ and ultimate CP divisible unital processes
satisfy the identity
\begin{equation}
\label{ultimate-lambda} \lambda_i(t) \lambda_j(t) = \lambda_k(t)\,.
\end{equation}

A general qubit unital semigroup evolution (up to unitary freedom)
takes the form
\begin{equation}
\label{eq:unital_generator}
\mathcal{L}[\varrho] = - \frac{1}{2}  \sum_{j=1}^{3}\Gamma_{j}
{\rm tr}[\sigma_{j} \varrho] \sigma_{j} = \frac{1}{2} \sum_{j=1}^3
\gamma_j \left( \sigma_j \varrho \sigma_j - \varrho \right),
\end{equation}

\noindent where $\Gamma_j^{-1}$ are experimentally measurable
timescales of decoherence processes and $\gamma_j$ are dissipator
rates given by formula
\begin{eqnarray}
\left(%
\begin{array}{c}
  \gamma_1 \\
  \gamma_2 \\
  \gamma_3 \\
\end{array}%
\right) = \frac{1}{2} \left(%
\begin{array}{ccc}
  -1 & 1 & 1 \\
  1 & -1 & 1 \\
  1 & 1 & -1 \\
\end{array}%
\right) \left(%
\begin{array}{c}
  \Gamma_1 \\
  \Gamma_2 \\
  \Gamma_3 \\
\end{array}%
\right).
\end{eqnarray}

\begin{figure}
\includegraphics[width=8.5cm]{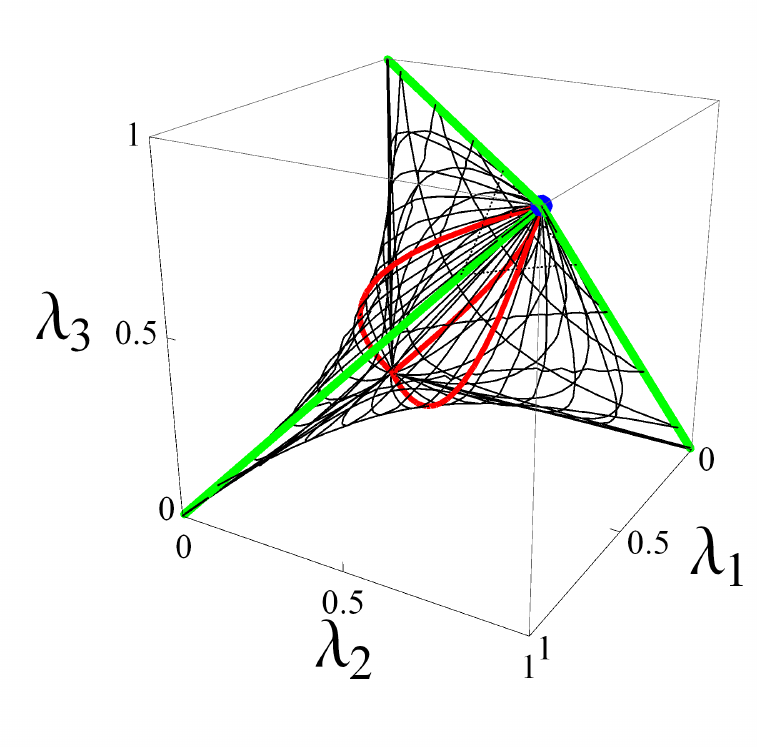}
\caption{\label{figure2} Ultimate CP divisible semigroups among
Pauli dynamical maps. Green lines correspond to pure dephasing
processes. Red curves correspond to generalized amplitude damping
processes with infinite temperature of the environment.}
\end{figure}

The conditions of ultimate CP divisibility on $\boldsymbol{\kappa}$
implies $\Gamma_i+\Gamma_j-\Gamma_k = \gamma_k =0$ (for some permutation
of indexes $i,j,k$). It follows that the generator for the ultimate
CP divisible Pauli semigroup contains at most two terms:
\begin{equation}
\label{dissipator-ultimate} \mathcal{L}[\varrho] =
\frac{\gamma_i}{2} \left( \sigma_i \varrho \sigma_i - \varrho
\right) + \frac{\gamma_j}{2} \left( \sigma_j \varrho \sigma_j -
\varrho \right),
\end{equation}

\noindent and the trajectory in the parameter space is $\lambda_i
= e^{-\gamma_j t}$, $\lambda_j = e^{-\gamma_i t}$, $\lambda_k =
e^{-(\gamma_i + \gamma_j) t}$. The class of time evolutions
for ultimate CP divisible Pauli semigroups is illustrated
in Fig.~\ref{figure2}. In the next section, we will provide
a physical realization of the generator \eqref{dissipator-ultimate}.


Physical examples of ultimate CP divisible processes include
\begin{itemize}
\item pure phase damping process, when $\lambda_i(t) = 1$,
  $\lambda_j (t) = \lambda_k (t) = e^{-\Gamma t}$ and corresponding to
  the choice of dissipation rates $\gamma_j=\gamma_k=0$
  (green lines in Fig.~\ref{figure2}).

\item generalized amplitude damping process with high-temperature
environment (\cite{nielsen-2000}, section 8.3.5), i.e. a
spontaneous decay with equal probabilities of energy absorption
and emission, when $\lambda_i(t) = \lambda_j(t) = e^{-\Gamma t}$
and $\lambda_k(t) = e^{-2\Gamma t}$ in Markov approximation
(\cite{breuer-petruccione-2002}, section 10.1).
\begin{equation}
\mathcal{L}[\varrho] = \Gamma \left( \sigma_+ \varrho \sigma_- +
\sigma_- \varrho \sigma_+ - \varrho \right),
\end{equation}

\noindent where $\sigma_{\pm} = \frac{1}{2} \left( \sigma_i \pm i
\sigma_j \right)$ are excitation creation and annihilation
operators. This process is illustrated as the bottom
red line in Fig.~\ref{figure2}
and corresponds to the choice of dissipation rates $\gamma_i=\gamma_j$
and $\gamma_k=0$.
\end{itemize}

Any Pauli channel $\Phi$ with parameters
$\lambda_1,\lambda_2,\lambda_3$ inside the body
determined by ultimate CP divisible processes in
Fig.~\ref{figure2} can be obtained as a result of some semigroup
dynamics with a particular generator $\mathcal{L}$ and time period
$t$, i.e. $\Phi = e^{\mathcal{L}t}$. Moreover,
even if the parameters of the
generator $\mathcal{L}$ in Eq.\eqref{eq:unital_generator} are
time-dependent but with positive decoherence rates (so-called
time-dependent Markovian dynamics~\cite{wolf-prl-2008}), then
achievable channels $\Phi_t$ still belong to the body in
Fig.~\ref{figure2}. Assigning equal weights to all Pauli channels,
the fraction of semigroup-achievable quantum channels equals
$\frac{V_{\rm body}}{V_{\rm tetrahedron}} = \frac{3}{32} =
9.375\%$, which is comparable with the numerical estimations of
general (non-unital) qubit semigroup-achievable channels (2\%) and
general (non-unital) qubit channels achievable by time-dependent
Markovian dynamics (17\%), Ref.~\cite{wolf-prl-2008}.

\section{Collision models of ultimate CP divisible semigroups}
\label{section-collision-semigroup}

\begin{figure}[b]
\includegraphics[width=8.5cm]{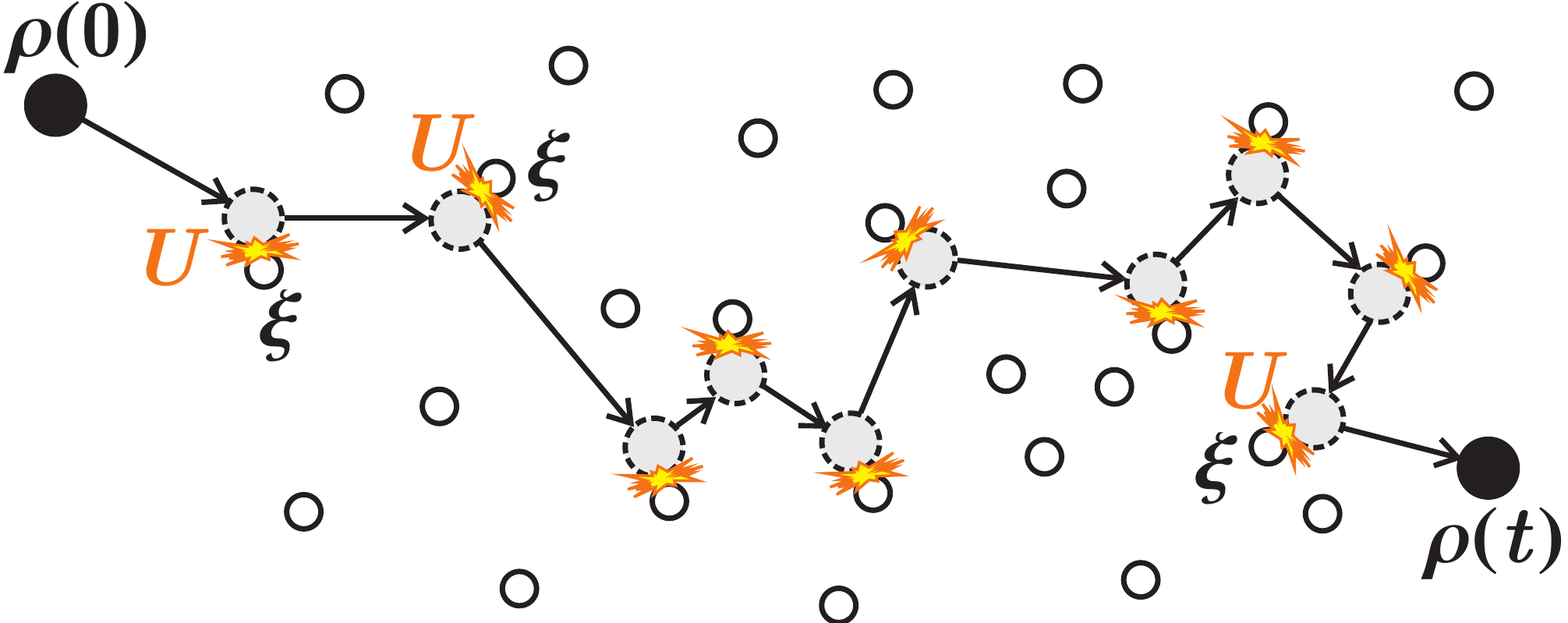}
\caption{\label{figure3} Physics of collision model.}
\end{figure}

\begin{figure}
\includegraphics[width=8.5cm]{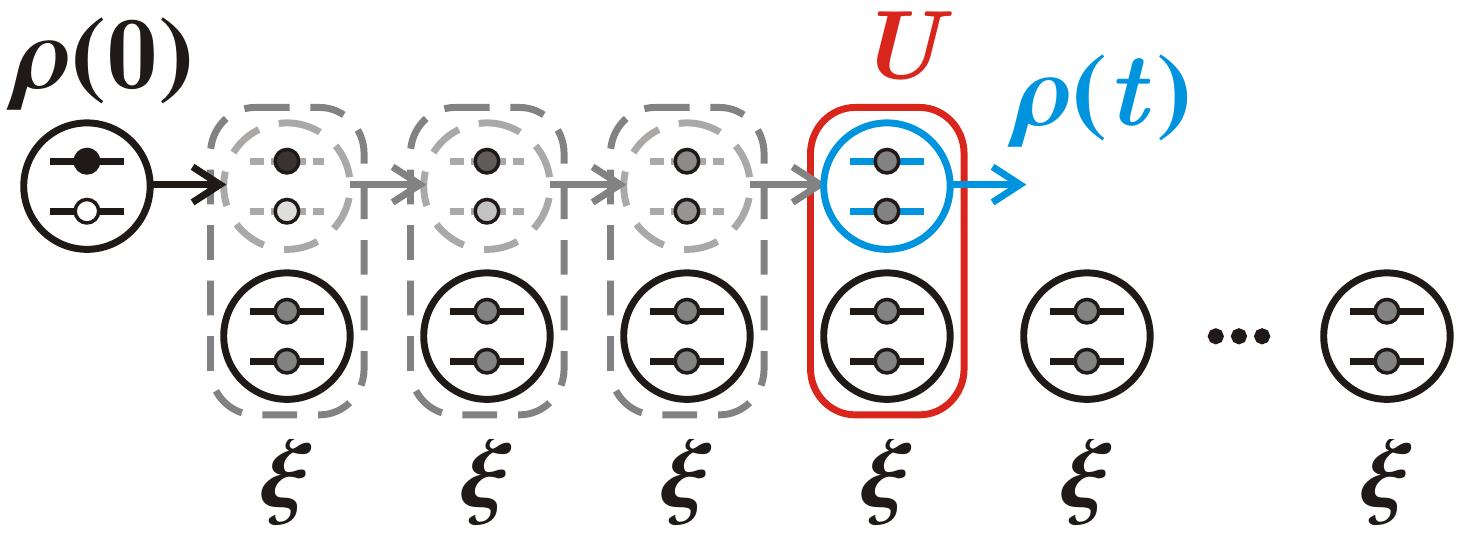}
\caption{\label{figure4} Collision model for Pauli dynamical maps
with ultimate CP divisible semigroup property.}
\end{figure}

Physically, the evolution $\frac{\partial}{\partial t} \varrho =
\mathcal{L}[\varrho]$ with dissipator \eqref{dissipator-ultimate}
is achievable as a result of sequential interactions of the system
qubit with environment qubits (collision model,
Fig.~\ref{figure3}). Let all environment qubits be in the same
state $\xi = \frac{1}{2} I$, Fig.~\ref{figure4}. The system qubit
and the $n$-th environment qubit interact pairwise during the time
period $\tau$, with the interaction Hamiltonian being
\begin{equation}
H_{\rm int} = \frac{1}{2} \left( g_1 \sigma_x \otimes \sigma_x +
g_2 \sigma_y \otimes \sigma_y \right).
\end{equation}

\noindent The system qubit and the $n$-th environment qubit
experience the unitary transformation
\begin{eqnarray}
&& U_{\tau} = \exp(-iH_{\rm int}\tau) \nonumber\\
&& = \cos\frac{g_1\tau}{2} \cos\frac{g_2\tau}{2} I \otimes I - i
\sin\frac{g_1\tau}{2} \cos\frac{g_2\tau}{2} \sigma_x \otimes
\sigma_x
\nonumber\\
&& - i \cos\frac{g_1\tau}{2} \sin\frac{g_2\tau}{2} \sigma_y
\otimes \sigma_y + \sin\frac{g_1\tau}{2}
\sin\frac{g_2\tau}{2} \sigma_z \otimes \sigma_z. \nonumber\\
\end{eqnarray}

\noindent As a result of such an interaction, the system state
$\varrho$ transforms as follows:
\begin{equation}
\varrho \longrightarrow \Phi_{\tau} [\varrho] = {\rm tr}_{n}
\left\{ U_{\tau} \left( \varrho \otimes \tfrac{1}{2} I \right)
U_{\tau}^{\dag} \right\},
\end{equation}

\noindent where ${\rm tr}_n$ denotes the partial trace over $n$-th
environment qubit. Some algebra yields the single interaction
elementary map $\Phi_{\tau}$, which is unital, does not depend on
$n$ and reads
\begin{eqnarray}
&& \Phi_{\tau} [\varrho] = \frac{1}{2} \Big( {\rm tr}[\varrho]
I + \cos(g_2\tau) {\rm tr}[\sigma_x \varrho] \sigma_x  \nonumber\\
&& + \cos(g_1\tau) {\rm tr}[\sigma_y \varrho] \sigma_y +
\cos(g_1\tau)\cos(g_2\tau) {\rm tr}[\sigma_z \varrho]
\sigma_z \Big). \nonumber\\
\end{eqnarray}

Since the system qubit always interacts with a fresh environmental
particle, after $\frac{t}{\tau}$ interactions we get the dynamical
map
\begin{equation}
\Phi_t = \left( \Phi_{\tau} \right)^{t/ \tau}
\end{equation}

\noindent with parameters $\lambda_1(t) = [\cos(g_2\tau)]^{t/
\tau}$, $\lambda_2(t) = [\cos(g_1\tau)]^{t/ \tau}$, and
$\lambda_3(t) = \lambda_1(t) \lambda_2(t)$.  In the stroboscopic
limit~\cite{giovannetti-2012,luchnikov-2017,lorenzo-2017} $\tau
\rightarrow 0$,  $g_1^2 \tau \rightarrow 2 \gamma_1$, $g_2^2 \tau
\rightarrow 2 \gamma_2$ we get the continuous dynamics
$\lambda_1(t) = e^{-\gamma_2 t}$, $\lambda_2(t) = e^{-\gamma_1
t}$, and $\lambda_3(t) = e^{-(\gamma_1+\gamma_2) t}$. Thus,
parameters $\lambda_1(t)$, $\lambda_2(t)$, and $\lambda_3(t)$
satisfy condition \eqref{ultimate-lambda} and the induced dynamics
is ultimate CP divisible.
This proves that ultimate CP divisible dynamics with the dissipator
\eqref{dissipator-ultimate} can be realized in the stroboscopic
limit of the collision model with the elementary pairwise
Hamiltonian $H = g_i \sigma_i \otimes \sigma_i + g_j \sigma_j
\otimes \sigma_j$, where the coefficients $g_i$ and $g_j$ satisfy
$\frac{g_i}{g_j} = \sqrt{\frac{\gamma_i}{\gamma_j}}$. Trajectories
of ultimate CP divisible Pauli semigroups are depicted in
Fig.~\ref{figure2}.

\section{Multiplicativity and additivity of generators in collision models}
\label{section-multiplicativity}

Any CP divisible process $\Phi_t$ can be realized stroboscopically
via a collision model with the arbitrary chosen precision. In
fact, since $\Theta_{t,t+s}$ is a valid dynamical map for all $t$
and $s$, its dilation (unitary operator $V_{t,t+s}$ and
environment state $\xi_{t,t+s}$) is continuous with respect to $t$
and $s$~\cite{werner-2008}. Fixing $s=\tau$, we get a sequence of
environment states
\begin{equation}
\xi_{0,\tau}, \xi_{\tau,2\tau}, \ldots, \xi_{(n-1)\tau,n\tau},
\ldots
\end{equation}

\noindent and a sequence of unitary operators acting on the system
and $n$-th environment particle
\begin{equation}
V_{0,\tau}, V_{\tau,2\tau}, \ldots, V_{(n-1)\tau,n\tau}, \ldots
\end{equation}

\noindent such that the dynamics $\Phi_t[\varrho]$ coincides with
the simulation $\Phi_{n\tau}^{\rm sim}[\varrho] = {\rm tr}_{\rm
env}[ V_{(n-1)\tau,n\tau} \cdots V_{\tau,2\tau} V_{0,\tau} (
\varrho \otimes \xi_{0,\tau} \otimes \xi_{\tau,2\tau} \otimes
\ldots \otimes \xi_{(n-1)\tau,n\tau} ) V_{0,\tau}^{\dag}
V_{\tau,2\tau}^{\dag} \cdots V_{(n-1)\tau,n\tau}^{\dag}]$ at time
moments $t= n\tau$. Thus, there exists a collision model with
factorized environment which simulates master equation
$\frac{\partial \varrho}{\partial t} = \mathcal{L}_t [\varrho]$
for the generator $\mathcal{L}_t = \dot{\Phi}_t \circ \Phi_t^{-1}$
if $\Phi_t$ is CP divisible.

Analogously, if we replace the generator $\mathcal{L}_t$ by
$\alpha \mathcal{L}_t$ with some positive $\alpha$, then the
resulting evolution is still CP divisible and can be realized
stroboscopically at the same time moments $t= n\tau$ (each
collision increments time by $\tau$) with a modified sequence of
environment states
\begin{equation}
\xi_{0,\alpha\tau}, \xi_{\tau,(1+\alpha)\tau}, \ldots,
\xi_{(n-1)\tau,(n - 1 + \alpha)\tau}, \ldots
\end{equation}

\noindent and a sequence of unitary operators acting on the system
and $n$-th environment particle
\begin{equation}
V_{0,\alpha\tau}, V_{\tau,(1+\alpha)\tau}, \ldots,
V_{(n-1)\tau,(n-1+\alpha)\tau}, \ldots
\end{equation}

Note that such an apparent construction of collision model for
multiplicative generator $\alpha\mathcal{L}_t$ is valid only if
the original process $\Phi_t$ is CP divisible. If this is not the
case, the modified master equation $\frac{\partial
\varrho}{\partial t} = \alpha \mathcal{L}_t [\varrho]$ may lead to
nonphysical solutions, with the example being presented in
Ref.~\cite{benatti-2017}.

\begin{figure}
\includegraphics[width=8.5cm]{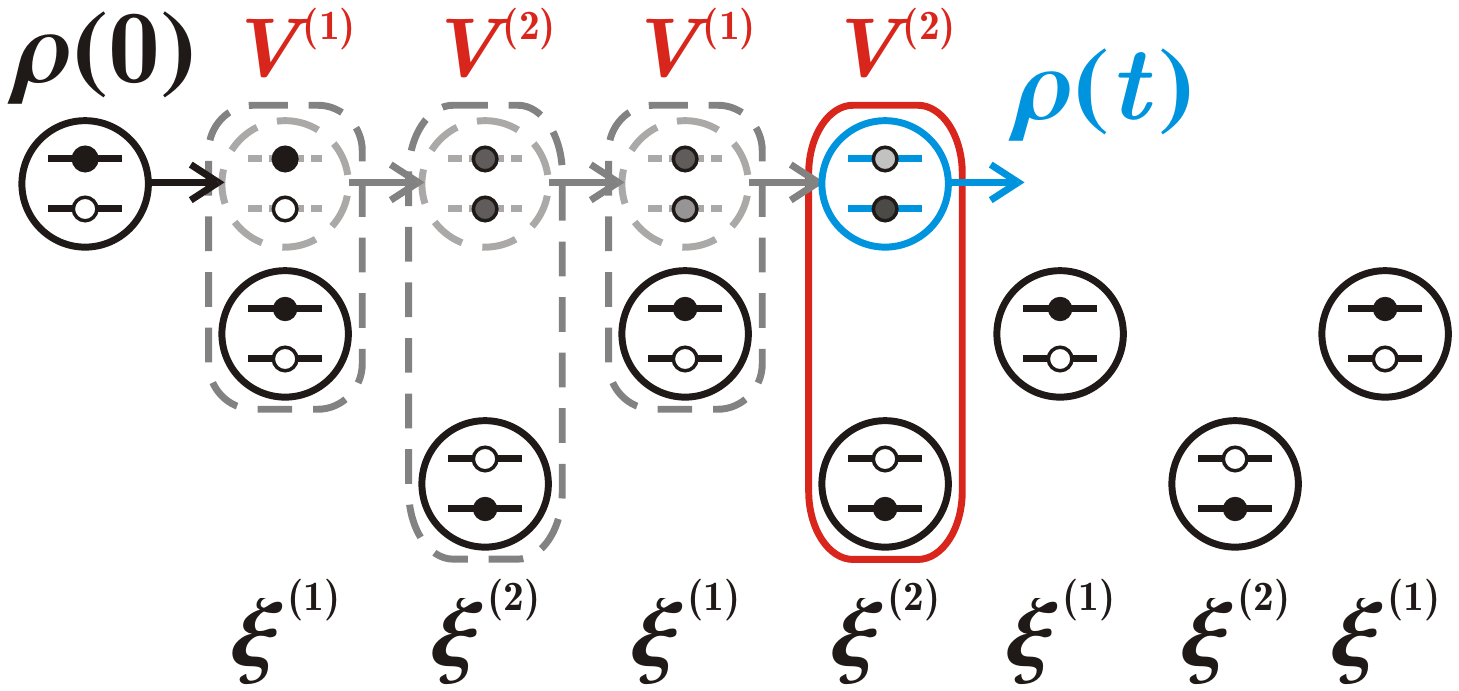}
\caption{\label{figure5} Simulation of generator $\frac{1}{2}
\left( \mathcal{L}_t^{(1)} + \mathcal{L}_t^{(2)} \right)$ for CP
divisible dynamical maps governed by master equations
$\frac{\partial \varrho}{\partial t} = \mathcal{L}_t^{(1)}
[\varrho]$ and $\frac{\partial \varrho}{\partial t} =
\mathcal{L}_t^{(2)} [\varrho]$.}
\end{figure}

Consider two CP divisible processes $\Phi_t^{(1)}$ and
$\Phi_t^{(2)}$ defined via master equations $\frac{\partial
\varrho}{\partial t} = \mathcal{L}_t^{(1)} [\varrho]$ and
$\frac{\partial \varrho}{\partial t} = \mathcal{L}_t^{(2)}
[\varrho]$, respectively. Each dynamical map $\Phi_t^{(i)}$ can be
simulated stroboscopically with a sequence of environment states
$\{ \xi_{(n-1)\tau,n\tau}^{(i)} \}$ and unitary operators $\{
V_{(n-1)\tau,n\tau}^{(i)} \}$, $i = 1,2$. If the system interacts
during time $\tau$ alternatively with particles from the first and
second sequences, i.e. with particles from the first environment
at odd collisions and with particles from the second environment
at even collisions (Fig.~\ref{figure5}), then the resulting
dynamics simulates the master equation $\frac{\partial
\varrho}{\partial t} = \frac{1}{2} \left( \mathcal{L}_t^{(1)} +
\mathcal{L}_t^{(2)} \right) [\varrho]$ at times $t= 2n\tau$. In a
more general physical situation, when the system interacts
independently with two types of environments, the effective
generator reads $p_1 \mathcal{L}_t^{(1)} + p_2
\mathcal{L}_t^{(2)}$, where $p_1$ and $p_2$ are the probabilities
of encountering a particle from the first and second environment,
respectively. Therefore, additivity of generators can be realized
in a stroboscopic model if those generators lead to CP divisible
dynamics. When the latter condition is violated, addition of
generators may also lead to nonphysical
solutions~\cite{kolodynski-2017}.

\section{Mixtures of CP divisible processes}
\label{section-mixtures}

Consider a dynamical map which is a mixture of CP divisible
processes:
\begin{equation}
\label{mixture-of-maps} \Phi_t = \sum_{m=1}^M p_m \Phi_t^{(m)},
\end{equation}

\noindent where $\{p_m\}$ are the probabilities with which CP
divisible dynamical maps $\{\Phi_t^{(m)}\}$ contribute to the map
$\Phi_t$, $p_m \geqslant 0$ and $\sum_{m=1}^M p_m = 1$. Note that
this situation is substantially different from the weighted sum of
generators since $\mathcal{L}_t = \dot{\Phi}_t \circ \Phi_t^{-1}
\neq \sum_{m=1}^M w_m \dot{\Phi}_t^{(m)} \circ
\left(\Phi_t^{(m)}\right)^{-1} = \sum_{m=1}^M w_m
\mathcal{L}_t^{(m)}$ in general.

\begin{figure}[b]
\includegraphics[width=8.5cm]{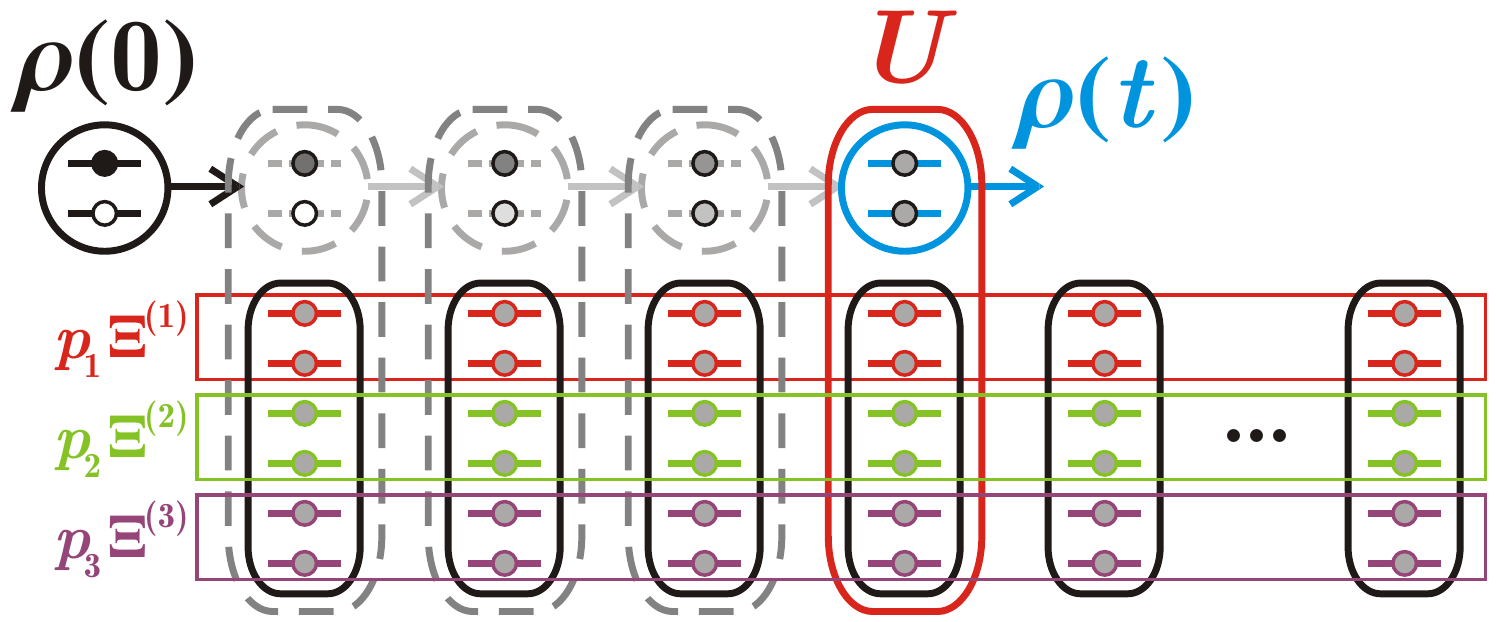}
\caption{\label{figure6} Collision model with correlated
environment, which realizes deterministic mixture of CP divisible
processes.}
\end{figure}

Surprisingly, even if all the processes $\Phi_t^{(m)}$ are CP
divisible, $\Phi_t$ can still be CP indivisible. The prominent
example is the mixture
\begin{equation}
\label{dephasing-mixture} \Phi_t^{\rm mix} = p_1 e^{\mathcal{L}_1
t} + p_2 e^{\mathcal{L}_2 t} + p_3 e^{\mathcal{L}_3 t}
\end{equation}

\noindent of purely dephasing maps $e^{\mathcal{L}_i t}$ with
$\mathcal{L}_i [\varrho] = \gamma (\sigma_i \varrho \sigma_i -
\varrho)$. In Ref.~\cite{megier-2016} the region of simplex
$(p_1,p_2,p_3)$ is found, for which $\Phi_t^{\rm mix}$ is not CP
divisible for all $t
> t_{\ast}$. If only one of probabilities $p_1,p_2,p_3$ equals zero, then $\Phi_t^{\rm
mix}$ is eternal CP indivisible.


In what follows we will design a collision model simulation of a general
mixture $\Phi_t = \sum_{m=1}^M p_m \Phi_t^{(m)}$ of CP divisible dynamical maps.
Suppose each $\Phi_t^{(m)}$ is realized by a collision model with
environment states $\xi_1^{(m)}$, $\xi_2^{(m)}$, $\ldots$,
$\xi_n^{(m)}$ and elementary unitary transformations $U_1^{(m)} =
\exp(-i H_1^{(m)} \tau)$, $U_2^{(m)} = \exp(-i H_2^{(m)} \tau)$,
$\ldots$, $U_n^{(m)} = \exp(-i H_n^{(m)} \tau)$. The whole
environment of $m$-th process reads $\Xi^{(m)} = \xi_1^{(m)}
\otimes \xi_2^{(m)} \otimes \cdots \otimes \xi_n^{(m)} =
\bigotimes\limits_{k=1}^n \xi_k^{(m)}$. In a probabilistic sense
the mixture can be realized as a mixture of collision models for each
individual $\Phi_t^{(m)}$, however,
such implementation is not ``operationally faithful'', because in each run
of the experiment a randomly chosen but different CP divisible process
$\Phi_t^{(m)}$ is realized. We will present an alternative realization of
such mixtures and design a collision model with a
\emph{correlated state of the environment} implementing the
desired mixture in each individual run of the experiment.

In particular,
consider the following initial state of the environment
\begin{eqnarray}
&& \Xi = \bigoplus\limits_{m=1}^M p_m \Xi^{(m)} = \left(
\begin{array}{cccc}
 p_1 \Xi^{(1)} &  &  &  \\
 & p_2 \Xi^{(2)} &  &  \\
 &  & \ddots &   \\
 &  &  &  p_M \Xi^{(M)}
\end{array}
 \right) \nonumber\\
 && = \left(
\begin{array}{cccc}
\blockmatrix{0.7in}{0.35in}{$p_1 \bigotimes\limits_{k=1}^n
\xi_k^{(1)}$} & \multicolumn{3}{|c}{}\\
\cline{1-2} \multicolumn{1}{c|}{} &
\blockmatrix{0.7in}{0.35in}{$p_2 \bigotimes\limits_{k=1}^n
\xi_k^{(2)}$} &\multicolumn{2}{|c}
{\raisebox{1.5ex}[0pt]{\parbox{12pt}{\Huge 0}}}\\
\cline{2-2} \multicolumn{2}{c}{}& \ddots &   \\
\cline{4-4} \multicolumn{3}{c|}
{\raisebox{-1.0ex}[0pt]{\parbox{12pt}{\Huge 0}}} &
\blockmatrix{0.7in}{0.35in}{$p_M \bigotimes\limits_{k=1}^n
\xi_k^{(M)}$}
\end{array}
 \right), \nonumber\\ \label{combined-environment}
\end{eqnarray}

\noindent which does not have the tensor product structure with
respect to collisions, i.e. $\Xi \neq \xi_1 \otimes \xi_2 \otimes
\cdots \otimes \xi_n$. Let us note that this state is correlated,
but not entangled. Also, note that the Hermitian operator
$H_k^{(m)}$ is an interaction Hamiltonian between the system and
the $k$-th particle of $m$-th environment, so $H_k^{(m)}$ acts
non-trivially on vectors in the subspace $\mathcal{H}_{\rm sys}
\otimes \mathcal{H}_k^{(m)}$ only. In other words, $H_k^{(m)}$
involves degrees of freedom of the system and the $k$-th particle
of $m$-th block of matrix \eqref{combined-environment}.
Consequently, $H_k^{(m)} H_{k}^{(m')} = 0$ if $m \neq m'$. The
combined Hamiltonian $H_k = \sum_{m=1}^M H_k^{(m)}$ generates the
unitary evolution operator
\begin{eqnarray}
&& U_k = \exp\left( -i H_k \tau \right) = \sum_{l=0}^{\infty}
\frac{(-i \tau)^l}{l!} (H_k)^l = \sum_{l=0}^{\infty} \frac{(-i \tau)^l}{l!} \nonumber\\
&& \times \sum_{m=1}^M \left( H_k^{(m)} \right)^l = \sum_{m=1}^M
\exp\left( -i H_k^{(m)}
\tau \right) = \sum_{m=1}^M U_k^{(m)}, \nonumber\\
\end{eqnarray}

\noindent where the support of $U_k^{(m)}=\exp\left( -i H_k^{(m)}
\tau \right)$ is $\mathcal{H}_{\rm sys} \otimes
\mathcal{H}_k^{(m)}$, so $U_k^{(m)} U_{k}^{(m')} = 0$ if $m \neq
m'$. Sequence of $n$ collisions results in the evolution operator
\begin{equation}
U_n \cdots U_2 U_1 = \sum_{m=1}^M U_n^{(m)} \cdots  U_2^{(m)}
U_1^{(m)},
\end{equation}

\noindent where $U_n^{(m)} \cdots  U_2^{(m)} U_1^{(m)}$ does not
vanish on vectors involving the system and the $m$-th block of
matrix \eqref{combined-environment}.

The dynamical map after $n$ collisions reads
\begin{eqnarray}
&& \Phi[\varrho] = {\rm tr}_{\rm env} \left[ U_n \cdots U_2 U_1 \
\varrho \otimes \Xi \ U_1^{\dag} U_2^{\dag} \cdots
U_n^{\dag} \right] \nonumber\\
&& = \sum_{m=1}^M p_m \, {\rm tr}_{\rm env} \left[ U_n^{(m)}
\cdots U_1^{(m)} \varrho \otimes \Xi^{(m)} U_1^{(m)\dag}
\cdots U_n^{(m)\dag} \right] \nonumber\\
&& = \sum_{m=1}^M p_m \Phi^{(m)}[\varrho]. \label{mixture-CP}
\end{eqnarray}

\noindent Therefore, the correlated environment
\eqref{combined-environment} enables realization of the mixture of
dynamical maps \eqref{mixture-of-maps}.

\begin{example}
Consider a mixture of pure dephasing qubit channels,
Eq.~\eqref{dephasing-mixture}, $M=3$. \textit{Deterministic}
collision model of such a dynamics is achieved with the
environment composed of $n$ 6-level systems, Fig.~\ref{figure6}.
The classically correlated state of $n$ environment particles is
\begin{equation}
\Xi = \left( p_1 \bigotimes\limits_{k=1}^n \frac{1}{2} I^{(1)}
\right) \bigoplus \left( p_2 \bigotimes\limits_{k=1}^n \frac{1}{2}
I^{(2)} \right) \bigoplus \left( p_3 \bigotimes\limits_{k=1}^n
\frac{1}{2} I^{(3)} \right),
\end{equation}

\noindent which assigns probability $p_1$ $(p_2,p_3)$ to the
occurrence of collision with the first (second, third) pair of
levels within the 6-level system.

Elementary unitary transformations $U_{k}^{(m)}$ coincide for all
collisions $k=1,\ldots,n$ and represent a generalization of a
controlled-unitary operation, where the system is a controlled
qubit, and $m$-th qubit within the triple serves as a controlling
qubit:
\begin{equation}
U_k^{(m)} =  e^{i g\tau \sigma_m} \otimes \ket{0_m}\bra{0_m} +
e^{-i g\tau \sigma_m} \otimes \ket{1_m}\bra{1_m}.
\end{equation}

In the stroboscopic
limit~\cite{giovannetti-2012,luchnikov-2017,lorenzo-2017} $\tau
\rightarrow 0$ and $g^2 \tau \rightarrow 2 \gamma$,
Eq.~\eqref{mixture-CP} leads to the dynamical map $\Phi_t^{\rm
mix} = p_1 e^{\mathcal{L}_1 t} + p_2 e^{\mathcal{L}_2 t} + p_3
e^{\mathcal{L}_3 t}$ with $\mathcal{L}_i [\varrho] = \gamma
(\sigma_i \varrho \sigma_i - \varrho)$.
\end{example}

\section{Eternal CP indivisibility}
\label{section-eternal}

Eternal CP indivisible dynamical maps $\Phi_t$ are those that are not CP
divisible for any time $t>0$. It was shown recently that eternal CP
indivisibility is quite a general property for spin-boson
systems~\cite{li-2017}. Known examples of eternal CP indivisible
Pauli dynamical maps include non-trivial convex combinations $p_i
\Phi_t^{{\rm pd} i} + p_j \Phi_t^{{\rm pd} j}$ of pure dephasing
processes $\Phi_t^{{\rm pd} i}$ and $\Phi_t^{{\rm pd} j}$ (in the
basis of eigenstates of operators $\sigma_i$ and $\sigma_j$,
respectively), $i,j = x,y,z$, $i \neq
j$~\cite{hall-2014,megier-2016}. In what follows, we extend this
one-parameter family (since $p_i + p_j =1$) to a wider class,
namely, a two-parameter family of eternal CP indivisible maps. The
underlying idea is to consider such smooth trajectories
$\boldsymbol{\lambda}(t)$ in the parameter space
$\lambda_1,\lambda_2,\lambda_3$ that do not belong to the
geometrical body in Fig.~\ref{figure2}. These trajectories are
beyond ultimate CP divisible processes, as a result
$\boldsymbol{\kappa}$-vector always points beyond the tetrahedron
in Fig.~\ref{figure1}b.

\begin{figure}
\includegraphics[width=8.5cm]{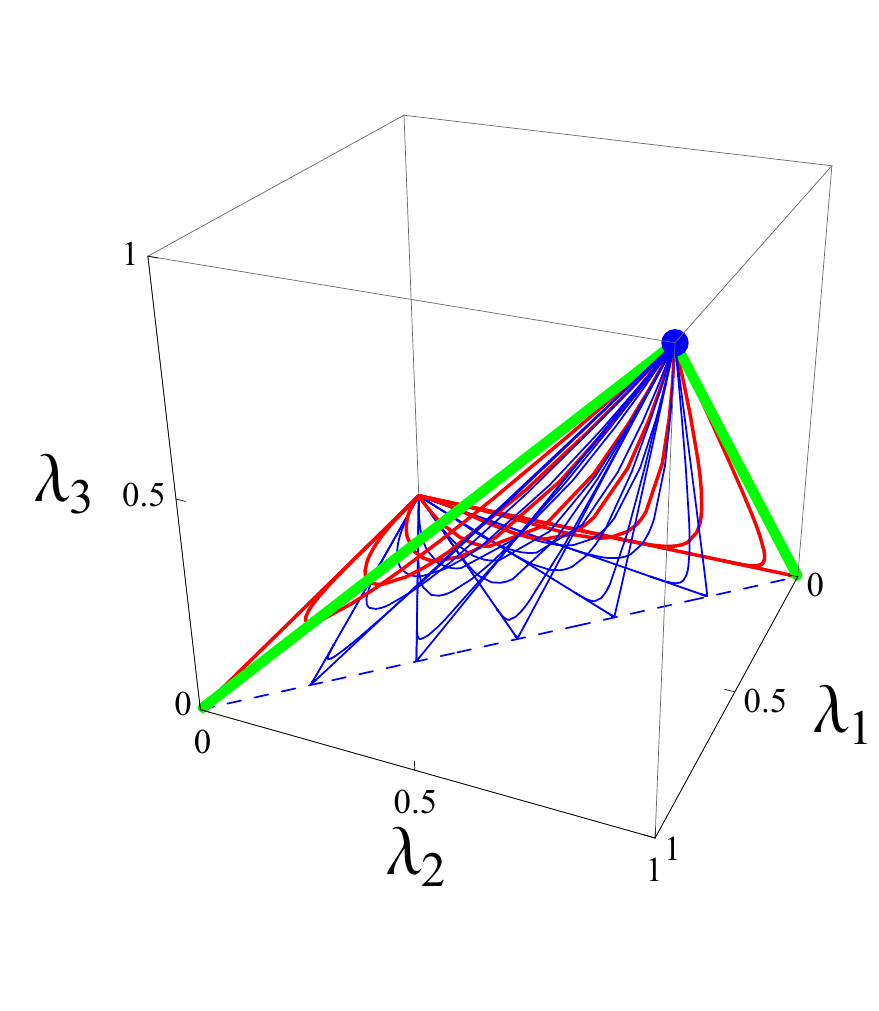}
\caption{\label{figure7} Trajectories of eternal CP indivisible
Pauli dynamical maps (blue lines), which are convex mixtures of
pure dephasing processes (green lines) and one of non-trivial
ultimate CP divisible maps (one of red lines).}
\end{figure}

To start with, focus on three ultimate CP divisible semigroup
processes:
\begin{itemize}

\item $\Phi_t^{(1)} = e^{\mathcal{L}^{(1)}t}$ with the dissipator
$\mathcal{L}^{(1)}[\varrho] = \frac{\gamma_i}{2} \left( \sigma_i
\varrho \sigma_i - \varrho \right) + \frac{\gamma_j}{2} \left(
\sigma_j \varrho \sigma_j - \varrho \right)$, $i \neq j$,
$\gamma_{i,j}
> 0$, which describes a ``skewed'' amplitude damping process
towards a completely mixed state $\frac{1}{2}I$ via contact with
high-temperature environment;

\item $\Phi_t^{(2)} = e^{\mathcal{L}^{(2)}t}$ with the dissipator
$\mathcal{L}^{(2)}[\varrho] = \frac{\gamma_i}{2} \left( \sigma_i
\varrho \sigma_i - \varrho \right)$, which is a pure phase damping
process in the basis of eigenstates of $\sigma_i$;

\item $\Phi_t^{(3)} = e^{\mathcal{L}^{(3)}t}$ with the dissipator
$\mathcal{L}^{(3)}[\varrho] = \frac{\gamma_j}{2} \left( \sigma_j
\varrho \sigma_j - \varrho \right)$, which is a pure phase damping
process in the basis of eigenstates of $\sigma_j$.

\end{itemize}

Let us demonstrate that any non-trivial mixture $\Phi_t = p_1
\Phi_t^{(1)} + p_2 \Phi_t^{(2)} + p_3 \Phi_t^{(3)}$ with
$p_{1,2,3} > 0$ is eternal CP indivisible. In fact, parameters of
the unital map $\Phi_t$ read
\begin{eqnarray}
\lambda_{i}(t) &=& (p_1 + p_3) e^{-\gamma_j t} + p_2, \\
\lambda_{j}(t) &=& (p_1 + p_2) e^{-\gamma_i t} + p_3, \\
\lambda_{k}(t) &=& p_1 e^{-(\gamma_i + \gamma_j)t} + p_2
e^{-\gamma_i t} + p_3 e^{-\gamma_j t}.
\end{eqnarray}

Calculation of the $\boldsymbol{\kappa}$-vector yields
\begin{eqnarray}
\kappa_{i}(t) &=& - \frac{\gamma_j(p_1 + p_3)}{p_1 + p_3 + p_2 e^{\gamma_j t}}  , \\
\kappa_{j}(t) &=& - \frac{\gamma_i(p_1 + p_2)}{p_1 + p_2 + p_3 e^{\gamma_i t}}, \\
\kappa_{k}(t) &=& - \frac{\gamma_i(p_1 + p_2 e^{\gamma_j t}) +
\gamma_j(p_1 + p_3 e^{\gamma_i t})}{p_1 + p_2 e^{\gamma_j t} + p_3
e^{\gamma_i t}}.
\end{eqnarray}

Since the inequalities
\begin{eqnarray}
\frac{p_1 + p_2 e^{\gamma_j t}}{p_1 + p_2 e^{\gamma_j t} + p_3
e^{\gamma_i t}} &>& \frac{p_1 + p_2}{p_1 + p_2 + p_3 e^{\gamma_i
t}}, \\
\frac{p_1 + p_3 e^{\gamma_i t}}{p_1 + p_2 e^{\gamma_j t} + p_3
e^{\gamma_i t}} &>& \frac{p_1 + p_3}{p_1 + p_3 + p_2 e^{\gamma_j
t}}
\end{eqnarray}

\noindent hold true for all $t > 0$, we conclude that $\kappa_i +
\kappa_j - \kappa_k > 0$ and one of inequalities
\eqref{CP-1}--\eqref{CP-3} is violated. Thus, $\Phi_t = p_1
\Phi_t^{(1)} + p_2 \Phi_t^{(2)} + p_3 \Phi_t^{(3)}$ is eternal CP
indivisible.

Thus, we have constructed a two-parameter family (since
$p_1+p_2+p_3=1$) of eternal CP indivisible processes as a mixture
of three ultimate CP divisible dynamical maps with clear physical
meaning. This family comprises the previously known examples as a
partial case when $p_1 = 0$. Corresponding trajectories in the
parameter space are depicted in Fig.~\ref{figure7}. Note, that the
constructed family is a mixture of CP divisible processes, so it
can be realized by a collision model developed in the previous
section.

\section{P divisibility}
\label{section-p-divisibility}

In this section, we review elementary operational
features of P divisible dynamical processes.

\subsection{Probability of confusion}

Consider a positive map $\Theta$~\footnote{Positive map is a linear
map that transforms positive semidefinite operators into positive
semidefinite ones. Hereafter, we assume the dimensions of the
input and the output spaces to be equal. Also, for the sake of
brevity, we will refer to positive trace preserving maps as
positive.}, then the quantum relative entropy $S(\varrho \|
\sigma) = {\rm tr}[\varrho(\ln \varrho - \ln \sigma)]$ is a
monotone under positive maps~\cite{muller-hermes-2017}, i.e.
\begin{equation}
S( \Theta[\varrho]  \| \Theta[\sigma] ) \leqslant S( \varrho \|
\sigma )
\end{equation}

\noindent for all density matrices $\varrho$ and $\sigma$. On the
other hand, quantum analogue of Sanov's theorem~\cite{vedral-1997}
states that the probability of confusing two quantum states
$\varrho$ and $\sigma$ after performing $n$ measurements on
$\sigma$ equals
\begin{equation}
P_n(\sigma \rightarrow \varrho) = e^{-n S(\varrho \| \sigma)}
\quad \text{if} \quad n \gg 1.
\end{equation}

Therefore, the probability of confusing two states $\varrho$ and
$\sigma$ monotonically increases in P divisible processes
[$S(\varrho \| \sigma)$ monotonically decreases].

\subsection{Distinguishability}

The trace distance $D(\varrho,\sigma) = \frac{1}{2} \| \varrho -
\sigma \|_1$ between qubit states $\varrho$ and $\sigma$ is a
monotone under qubit positive maps $\Theta$ too, i.e.
\begin{equation}
D(\Theta[\varrho], \Theta[\sigma]) \leqslant D(\varrho,\sigma).
\end{equation}

On the other hand, the trace distance quantifies the probability
of successful discrimination of quantum states $\varrho$ and
$\sigma$ in a single-shot measurement. For P divisible processes
this probability monotonically decreases.

\subsection{Classical capacity}

If the process $\Phi_t$ is unital, then the map $\Theta_{t,t+s}$
is also unital. Classical capacity $C$ of a qubit unital channel
reads $C(\Phi_t) = 1 - h_2\left[ \frac{1}{2}\big(1 -
\max(|\lambda_1(t)|,|\lambda_2(t)|,|\lambda_3(t)|)\big) \right]$,
where $h_2(x) = - x {\rm log}_2 x - (1-x) {\rm log}_2 (1-x)$. It
is not hard to see, that all $|\lambda_i(t)|$, $i=1,2,3$,
monotonically decrease if $\Theta_{t,t+s}$ is positive for all
$t,s$. Therefore, if the qubit unital process $\Phi_t$ is P
divisible, then its classical capacity $C(\Phi_t)$ monotonically
decreases with time $t$.

\subsection{Separability}

If a positive map $\Theta$ is applied to a part of separable state
$R = \sum_i \pi_i \varrho_i \otimes \sigma_i$, $\pi_i \geqslant
0$, then its separability is preserved since $(\Theta \otimes {\rm
Id}) [R] = \sum_i \pi_i \Theta[\varrho_i] \otimes \sigma_i$ is a
valid density operator. Thus, if the process $\Phi_t$ is P
divisible, then its action on a part of a composite system cannot
result in the revival of entanglement.

Suppose that by time $t=t_{\rm EB}$ the process $\Phi_t$ becomes
entanglement breaking~\cite{holevo-1998,horodecki-2003}, i.e.
$\Phi_{t_{\rm EB}}$ is effectively a measure-and-prepare procedure
(quantum-classical-quantum channel) of the Holevo form
$\Phi_{t_{\rm EB}} [\varrho] = \sum_{k} {\rm tr}[\varrho E_k]
\varrho_k$, where $\{E_k\}$ is a positive operator-valued measure.
If $\Phi_t$ acts on a part of a composite system (initially in the
state $R_0$), then $(\Phi_{t_{\rm EB}} \otimes {\rm Id})[R_0]$ is
separable and the further P divisible dynamics leaves this state
separable.

Suppose the channel $\Phi_t \otimes \Phi_t$ becomes
entanglement-annihilating~\cite{moravcikova-ziman-2010,filippov-rybar-ziman-2012}
by time $t=t_{\rm EA}$, and $\Phi_{t}$ is P divisible for $t >
t_{\rm EA}$. Then $(\Phi_{t_{\rm EA}} \otimes \Phi_{t_{\rm
EA}})[R_0]$ is separable and $(\Phi_{t} \otimes \Phi_{t})[R_0]$
remains separable for $t
> t_{\rm EA}$.

For instance, the Pauli channel $\Phi$ with parameters
$\lambda_1,\lambda_2,\lambda_3$ results in
entanglement-annihilating channel $\Phi \otimes \Phi$ if and only
if $\lambda_1^2 + \lambda_2^2 + \lambda_3^2 \leqslant
1$~\cite{filippov-rybar-ziman-2012}. The process $\Phi_t = p_1
e^{\mathcal{L}_1 t} + p_2 e^{\mathcal{L}_2 t} + p_3
e^{\mathcal{L}_3 t}$ with dissipators $\mathcal{L}_i [\varrho] =
\gamma (\sigma_i \varrho \sigma_i - \varrho)$ becomes entanglement
annihilating if
\begin{equation}
p_1^2 + p_2^2 + p_3^2 = \frac{1 - e^{-\gamma t_{\rm EA}} - e^{- 2
\gamma t_{\rm EA}}}{1 - e^{-\gamma t_{\rm EA}} + e^{- 2 \gamma
t_{\rm EA}}}.
\end{equation}

Positive divisibility of the map $\Phi_t$ guarantees separability
of $(\Phi_t \otimes \Phi_t)[R_0]$ for all $t > t_{\rm EA}$.

\subsection{Tensor power}

Clearly, a map $\Theta \otimes \Theta$ can be non-positive even if
$\Theta$ is
positive~\cite{muller-hermes-2016,filippov-magadov-2017}. Thus,
even if $\Phi_t$ is P divisible, $\Phi_t \otimes \Phi_t$ can still
be P indivisible. However, if $\Phi_t \otimes \Phi_t$ is P
divisible then $\Phi_t$ is CP divisible~\cite{benatti-2017}.

\section{Collision models for P indivisible dynamical maps}
\label{section-p-indivisibility-all}

P indivisible (essentially non-Markovian) dynamical maps $\Phi_t$
can exhibit properties opposite to those described in the previous
section, namely, the probability of confusion of two states,
distinguishability of states, and classical capacity can be
non-monotonic functions of time. In following subsections, we
construct collision models of specific and general P indivisible
processes and present an example of the dynamical map, which
monotonically shrinks the volume of accessible states but is not P
divisible.

\subsection{Essentially non-Markovian dephasing process}
\label{section-p-indivisibility}

\begin{figure}
\includegraphics[width=8.5cm]{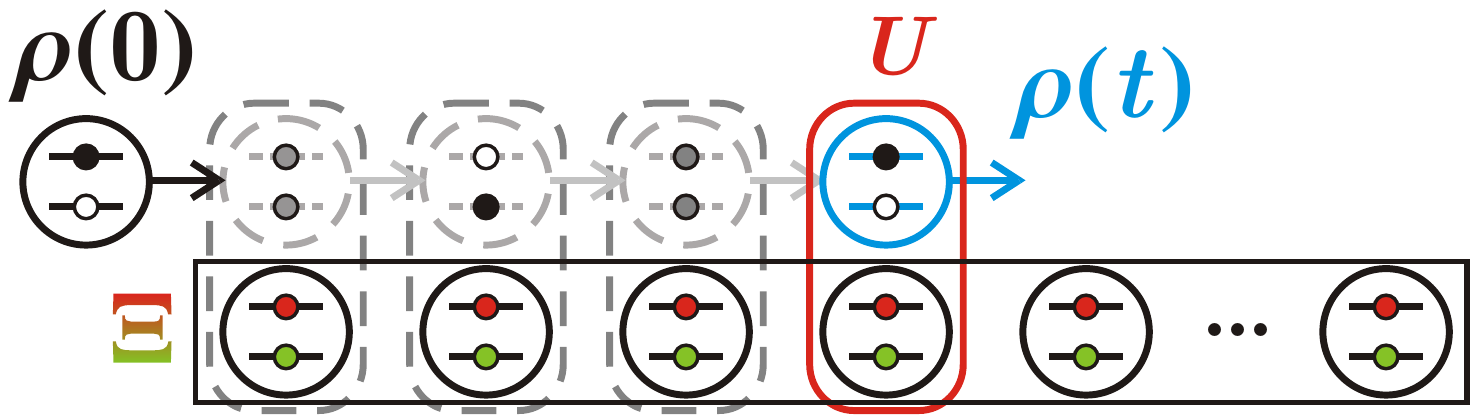}
\caption{\label{figure8} Essentially non-Markovian (P indivisible)
dynamics in collision model with correlated environment.
Correlations are encoded in color: either all environment qubits
are excited, or they are all in ground state.}
\end{figure}

As an example of P indivisible dynamics, consider a correlated
environment of $n$ qubits (Fig.~\ref{figure8}) in the state
\begin{equation}
\Xi = \frac{1}{2} \left( \ket{0^{\otimes n}} \bra{0^{\otimes n}} +
\ket{1^{\otimes n}} \bra{1^{\otimes n}} \right).
\end{equation}

The elementary unitary transformation $U_k$ describes the
evolution of the system and $k$-th environment qubit. Suppose $U_k
= e^{i g \tau \sigma_z} \otimes \ket{0}_k\bra{0} + e^{-i g \tau
\sigma_z} \otimes \ket{1}_k\bra{1}$, then after $n =
\frac{t}{\tau}$ collisions we get
\begin{eqnarray}
&& \Phi_t^z[\varrho] = {\rm tr}_{\rm env} \left[ U_n \cdots U_2
U_1 \ \varrho \otimes \Xi \ U_1^{\dag} U_2^{\dag} \cdots
U_n^{\dag} \right] \nonumber\\
&& = \cos^2(n g \tau) \varrho + \sin^2(n g \tau)
\sigma_z \varrho \sigma_z \nonumber\\
&& = \cos^2(g t) \varrho + \sin^2(g t) \sigma_z \varrho \sigma_z.
\label{P-indivisible-z}
\end{eqnarray}

Clearly, the resulting dephasing dynamics $\Phi_t$ is P divisible
if $0 < g t < \frac{\pi}{4}$ and P indivisible if $\frac{\pi}{4} <
g t < \frac{\pi}{2}$. Then the periods of P divisibility and P
indivisibility alternate. The information about the initial system
state $\varrho$ is stored in the environment when the process is P
divisible, and the back-flow of information occurs when the process
is P indivisible.

\subsection{Mixture of essentially non-Markovian dephasing processes}

Similarly to the previous subsection, P indivisible dephasing
processes $\Phi_t^x$ and $\Phi_t^y$ along $x$ and $y$ axes of the
Bloch ball can be achieved by collision models. The mixture
$\Phi_t = p_1 \Phi_t^x + p_2 \Phi_t^y + p_3 \Phi_t^z$ can be
realized with a correlated environment made of $n$ 6-level systems
in the state
\begin{equation}
\Xi = \sum_{m=1,2,3} \frac{p_m}{2} \left( \ket{0_m^{\otimes n}}
\bra{0_m^{\otimes n}} + \ket{1_m^{\otimes n}} \bra{1_m^{\otimes
n}} \right),
\end{equation}

\noindent where $m$ labels pairs of levels (effective qubit states
$\ket{0_m}$ and $\ket{1_m}$), and elementary unitary
transformations
\begin{equation}
U_k = \sum_{m = 1,2,3} e^{i g \tau \sigma_m} \otimes
\ket{0_m}_k\bra{0_m} + e^{-i g \tau \sigma_m} \otimes
\ket{1_m}_k\bra{1_m}.
\end{equation}

Pictorial representation of the resulting dynamical map
\begin{eqnarray}
\Phi_t &=& p_1 \Phi_t^x + p_2 \Phi_t^y + p_3 \Phi_t^z \nonumber\\
&=& \cos^2(\alpha t) \varrho + \sin^2(\alpha t) \sum_{m=1}^3 p_m
\sigma_m \varrho \sigma_m
\end{eqnarray}

\noindent is a straight line in the parameter space
$\lambda_1,\lambda_2,\lambda_3$.

For instance, in the case $p_1 = p_2 = p_3 = \frac{1}{3}$, we
obtain a depolarizing map $\mathcal{D}_{p(t)}[\varrho] = p(t)
\varrho + \big( 1-p(t) \big){\rm tr}[\varrho] \frac{1}{2} I$ with
$p(t) = \frac{1}{3}(1 + 2 \cos 2gt)$. In such a process, the Bloch
ball gradually shrinks to a point, then extends in inverted form
unless its radius equals $\frac{1}{3}$ (the best approximation of
universal NOT operation), and then the process goes in opposite
direction until the region of accessible states occupies the whole
Bloch ball again, after that the process continuous from the very
beginning.

\subsection{Arbitrary pure dephasing process}
\label{section-arbitrary-dephasing}

In subsection~\ref{section-p-indivisibility}, we considered a
essentially non-Markovian pure dephasing process with the
coherence function $\cos (2gt)$. In this subsection, we construct
a collision model which results in a pure dephasing process with
the arbitrary continuous real coherence function $f(t)$ that is
bounded ($|f(t)| \leqslant 1$) and $f(0)=1$.

We start with a dephasing process in the basis of eigenvectors of
$\sigma_z$, i.e. the density matrix transformation
\begin{equation}
\label{dephasing-arbitrary} \left(%
\begin{array}{cc}
  \varrho_{11} & \varrho_{12} \\
  \varrho_{21} & \varrho_{22} \\
\end{array}%
\right) \rightarrow \left(%
\begin{array}{cc}
  \varrho_{11} & f(t) \varrho_{12} \\
  f(t) \varrho_{21} & \varrho_{22} \\
\end{array}%
\right),
\end{equation}

\noindent which corresponds to a trajectory
$\boldsymbol{\lambda}(t) = \big( f(t),f(t),1 \big)$ in the
parameter space.

Consider a correlated environment in the state
\begin{eqnarray}
 \Xi & = & \frac{1}{2} \Big( \ket{i_1} \bra{i_1} \otimes \ket{i_2}
\bra{i_2} \otimes \cdots \otimes \ket{i_n} \bra{i_n} \otimes
\cdots \nonumber\\
&& +  \ket{\overline{i_1}} \bra{\overline{i_1}} \otimes
\ket{\overline{i_2}} \bra{\overline{i_2}} \otimes \cdots \otimes
\ket{\overline{i_n}} \bra{\overline{i_n}} \otimes \cdots \Big)
\nonumber\\
& = & \frac{1}{2} \bigotimes_{k} \ket{i_k} \bra{i_k} + \frac{1}{2}
\bigotimes_{k} \ket{\overline{i_k}} \bra{\overline{i_k}},
\label{Xi-arbitrary-z}
\end{eqnarray}

\noindent where either $i_k = 0$ and $\overline{i_k} = 1$, or $i_k
= 1$ and $\overline{i_k} = 0$. Elementary unitary transformations
$U_k = e^{i g \tau \sigma_z} \otimes \ket{0}_k\bra{0} + e^{-i g
\tau \sigma_z} \otimes \ket{1}_k\bra{1}$ result in the following
dynamical map after $n = \frac{t}{\tau}$ collisions:
\begin{eqnarray}
&& \!\!\!\!\!\!\! \Phi_t[\varrho] = {\rm tr}_{\rm env} \left[ U_n
\cdots U_2 U_1 \ \varrho \otimes \Xi \ U_1^{\dag} U_2^{\dag}
\cdots
U_n^{\dag} \right] \nonumber\\
&& \!\!\!\!\!\!\! = \cos^2 \{ [n_0(t) - n_1(t)] g \tau \} \varrho
+ \sin^2 \{
[n_0(t) - n_1(t)] g \tau \} \sigma_z \varrho \sigma_z , \nonumber\\
\label{arbitrary-z}
\end{eqnarray}

\noindent where $n_0(t) = \sum_{k=1}^n \delta_{i_k,0}$ and $n_1(t)
= \sum_{k=1}^n \delta_{i_k,1} =  n - n_0$. Apparently, $[n_0(t) -
n_1(t)] \tau = 2 n_0(t) \tau - t$ and
\begin{equation}
\label{f-function} f(t) = \cos\{ 2 g [2 n_0(t) \tau - t] \}.
\end{equation}

\noindent Therefore, to get the desired dynamics one needs to
arrange the number $n_0(t)$ of $0$'s in indices $i_k$ of
environment state \eqref{Xi-arbitrary-z} in accordance with the
formula
\begin{equation}
\label{n-0} n_0(t) \tau = \frac{\arccos f(t)}{4 g} + \frac{t}{2}.
\end{equation}

\noindent In the usual continuous limit $\tau \rightarrow 0$, $g
\tau \rightarrow {\rm const}$, the left hand side of
Eq.~\eqref{n-0} has the meaning of the integral $n_0(t) \tau =
\int_0^t w_0(t')dt'$, where $w_0(t)$ is the probability of
encountering $0$ at every collision in the first line of the
environment state \eqref{Xi-arbitrary-z}. Finally,
\begin{equation}
w_0(t) = - \frac{f'(t)}{4 g \sqrt{1 - f^2(t)}} + \frac{1}{2}.
\end{equation}

If $f'(t) = 0$ when $f(t) = 1$, then the right hand side can be
made non-negative and bounded from above by 1 for sufficiently
large $g$.
If $f'(t) \neq 0$ when $f(t) = 1$ (as it takes place, e.g., in
Markov approximation), one has to resort to the stroboscopic limit
and replace $g$ by $\frac{g}{\sqrt{\tau}}$, which enables to meet
the requirement $0 \leqslant w_0(t) \leqslant 1$.
Similarly, one can construct the processes of arbitrary dephasing
in the bases of eigenstates of operators $\sigma_x$ and
$\sigma_y$.

\subsection{Arbitrary Pauli dynamical maps}
\label{section-arbitrary}

In this subsection, we construct a collision model which is able
to reproduce any dynamics $\boldsymbol{\lambda}(t)$ satisfying the
condition of complete positivity of the corresponding Pauli dynamical map
$\Phi_t$. In other words, given a trajectory in the parameter
space (Fig.~\ref{figure1}a), we construct a collision model
leading to such a trajectory.

The requirement of complete positivity is automatically fulfilled
if the functions $q_j(t)$ defined through
\begin{equation}
\label{q-lambda}
\left(%
\begin{array}{c}
  q_0(t) \\
  q_1(t) \\
  q_2(t) \\
  q_3(t) \\
\end{array}%
\right) = \frac{1}{4} \left(%
\begin{array}{cccc}
  1 & 1 & 1 & 1 \\
  1 & 1 & -1 & -1 \\
  1 & -1 & 1 & -1 \\
  1 & -1 & -1 & 1 \\
\end{array}%
\right) \left(%
\begin{array}{c}
  1 \\
  \lambda_1(t) \\
  \lambda_2(t) \\
  \lambda_3(t) \\
\end{array}%
\right)
\end{equation}

\noindent satisfy $q_j(t) \geqslant 0$ for all $j=0,1,2,3$.

To get an arbitrary Pauli dynamical map with non-negative
functions $q_0(t),q_1(t),q_2(t),q_3(t)$, one needs to combine
three (essentially non-Markovian) dephasing processes considered
in the previous section. This is achieved with the environment
composed of 3 types of qubits. Denote these types $x,y,z$, then
$k$-th collision of system qubit with $m$-th type of environment
qubits is described by the elementary unitary transformation
\begin{equation}
\label{elementary-arbitrary} U_k^{(m)} = e^{i g \tau \sigma_m}
\otimes \ket{0}_k\bra{0} + e^{-i g \tau \sigma_m} \otimes
\ket{1}_k\bra{1}.
\end{equation}

\begin{figure}
\includegraphics[width=8.5cm]{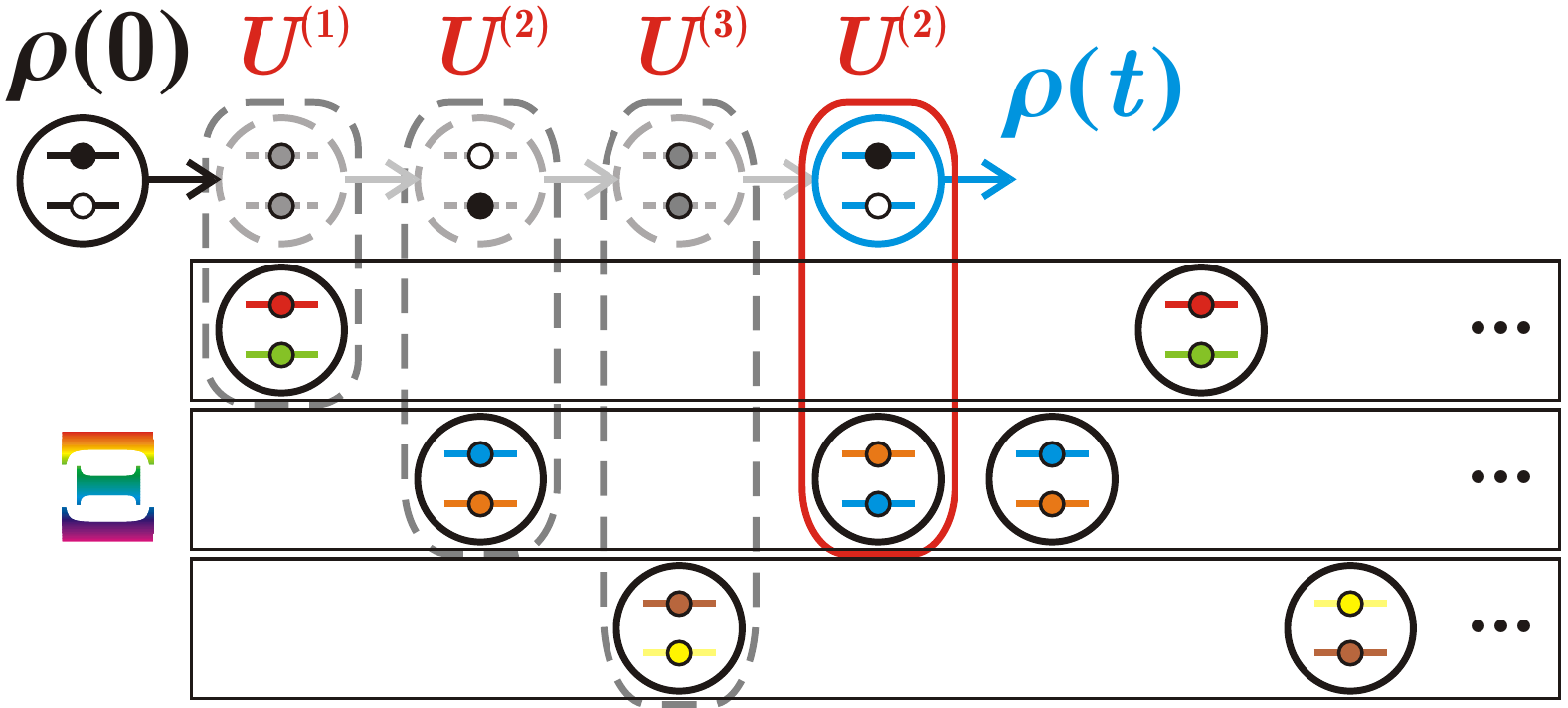}
\caption{\label{figure9} Collision model simulating essentially
non-Markovian Pauli dynamical maps.}
\end{figure}

\noindent Qubits of the same kind are correlated, so the total
environment state reads
\begin{equation}
\label{Xi-arbitrary} \Xi = \bigotimes_{m=x,y,z} \left( \frac{1}{2}
\bigotimes_{k \in \{ k_m \}} \ket{i_{k}}\bra{i_{k}} + \frac{1}{2}
\bigotimes_{k \in \{ k_m \}}
\ket{\overline{i_k}}\bra{\overline{i_k}} \right),
\end{equation}

\noindent where $\{k_x\}$, $\{k_y\}$, $\{k_z\}$ are subsequences
of collision numbers $k \in \mathbb{N}$ such that
$\{k_m\}\cap\{k_{m'}\} = \emptyset$ if $m \neq m'$ and $\{k_x\}
\cup \{k_y\} \cup \{k_z\} = \mathbb{N}$. Physics of such
collisions is depicted in Fig.~\ref{figure9}.

The resulting map is
\begin{eqnarray}
\Phi_t[\varrho] = {\rm tr}_{\rm env} \Big[ U_n\cdots U_1
(\varrho \otimes \Xi) \ U_1^\dagger\cdots U_n^\dagger\Big]\,,
\end{eqnarray}

\noindent where $U_k = U_{k}^{(m: \, k_m = k)}$. The intermediate
map $\Theta_{t,t+\tau} = \Phi_{t+\tau} \circ \Phi_t^{-1}$ between
collisions realizes one of the infinitesimal maps
$\Theta_{t,t+\tau}^{(x)}$, $\Theta_{t,t+\tau}^{(y)}$, and
$\Theta_{t,t+ \tau}^{(z)}$. Collision with $k_m$-th particle
results in the map $\Theta_{t,t+\tau}^{(m)}$. Clearly, for a fixed
$m$ the product $\prod_{k_m} \Theta_{k_m\tau,(k_m+1)\tau}^{(m)} =
\Phi_t^{(m)}$ is nothing else but the dephasing map in the
eigenbasis of operator $\sigma_m$ with dephasing function $f_m(t)$
given by a modification of Eq.~\eqref{f-function}:
\begin{equation}
\label{f-function-modified} f_m(t) = \cos \left\{ 2 \left[
n_0^{(m)} - n_1^{(m)} \right] g \tau \right\},
\end{equation}

\noindent where $n_0^{(m)} = \sum_{k_m \leqslant n}
\delta_{i_{k_m},0}$ and $n_1^{(m)} = \sum_{k_m \leqslant n}
\delta_{i_{k_m},1}$. All physical functions $f_m(t)$, $m=1,2,3$,
can be realized in the usual continuous or stroboscopic limit as
it was demonstrated for a single dephasing map. Then a sequence of
collisions with different types of qubits during a short time $dt$
($dt \gg \tau$) results in the product
\begin{equation}
\Theta_{t,t+d t} = \Theta_{t,t+d t}^{(x)} \Theta_{t,t+d t}^{(y)}
\Theta_{t,t+d t}^{(z)}.
\end{equation}

Note that all $\Theta_{t,t+d t}^{(m)}$ commute. Consequently, the
parameters $\lambda_1(t)$, $\lambda_2(t)$, and $\lambda_3(t)$ of
the map $\Phi_t$ satisfy differential equations $\lambda_1'(t) =
f_2'(t) + f_3'(t)$, $\lambda_2'(t) = f_1'(t) + f_3'(t)$,
$\lambda_3'(t) = f_1'(t) + f_2'(t)$, from which it follows that
$\lambda_1(t) = f_2(t) + f_3(t) -1$, $\lambda_2(t) = f_1(t) +
f_3(t) -1$, $\lambda_3(t) = f_1(t) + f_2(t) -1$. Finally, using
Eq.~\eqref{q-lambda}, we find the explicit form of the functions
$f_m(t)= 1 - 2 q_m(t)$.

The algorithm for producing arbitrary dynamics
$\boldsymbol{\lambda}(t)$ in parameter space is the following.
Calculate $q_m(t)$ by Eq.~\eqref{q-lambda} and $f_m(t)= 1 - 2
q_m(t)$. For each $m$ distribute $0$'s and $1$'s in accordance
with formula~\eqref{f-function-modified}. Create the correlated
state~\eqref{Xi-arbitrary} with corresponding distributions of
$0$'s and $1$'s in $m$'th branch. Let the system qubit interact
with environment qubits of type $m$ according to the elementary
evolution operator~\eqref{elementary-arbitrary}.

\subsection{Dynamical maps shrinking the volume of accessible states}
\label{section-volume}

\begin{figure}[t]
\includegraphics[width=8.5cm]{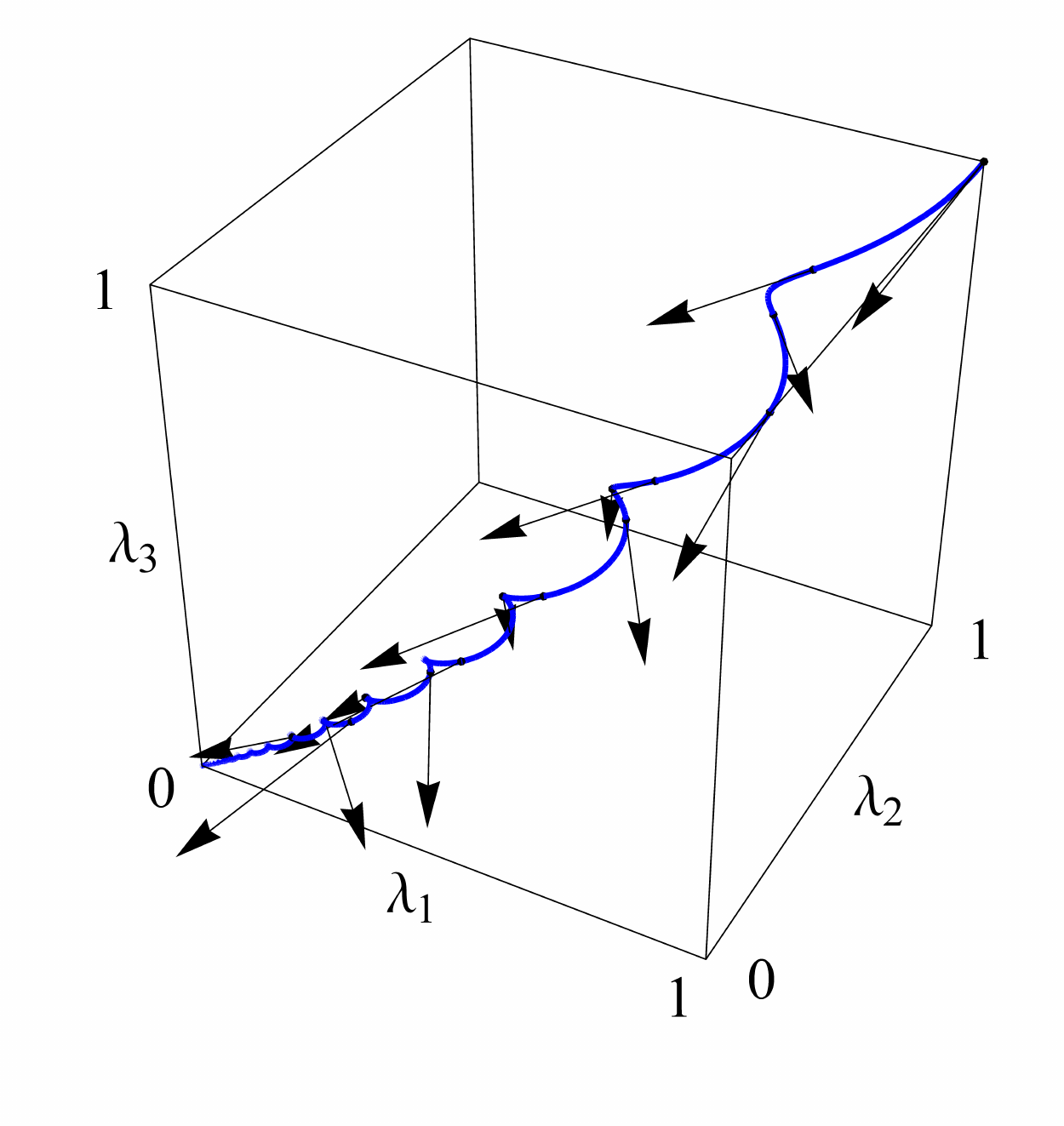}
\caption{\label{figure10} Trajectory of the Pauli dynamical map,
which is not P divisible but monotonically shrinks the volume of
accessible states. Arrows represent directions of the
$\boldsymbol{\kappa}$-vector at particular time moments.}
\end{figure}

One more approach to characterization of non-Markovianity is based
on the quantification of the volume of accessible
states~\cite{lorenzo-2013}. Using the metric induced by
Hilbert-Schmidt distance for qubit states, the volume of
accessible states of a qubit dynamical map $\Phi_t$ is simply the
volume of the ellipsoid in the Bloch ball picture, which
corresponds to the domain of $\Phi_t$. For Pauli dynamical maps
$\Phi_t$ of the form \eqref{unital-process} the volume of
accessible states $V(t) = | \lambda_1(t) \lambda_2(t) \lambda_3(t)
|$. A process $\Phi_t$ monotonically shrinks the volume of
accessible states if any intermediate map $\Theta_{t,t+s}$ does
so.

Let us present an example of the Pauli dynamical map $\Phi_t$
which monotonically shrinks the volume of accessible states but is
not P divisible. Let $\lambda_1(t) = e^{-2t}(1-\frac{1}{10}[1-\cos
40t])$, $\lambda_2(t) = e^{-2t}(1-\frac{1}{10}\sin 40t)$,
$\lambda_3(t) = e^{-4t}$, then $\Phi_t$ is a physical process
indeed since all $q_i(t) \geqslant 0$ and can be realized via
collision model with correlated environment (see the preceding
subsection). It is not hard to see that the volume of accessible
states $V(t)$ monotonically decreases, whereas both $\lambda_1(t)$
and $\lambda_2(t)$ are not monotonic, i.e. the process is not P
divisible (see the trajectory and the corresponding
$\boldsymbol{\kappa}$-vector in Fig.~\ref{figure10}).

\section{Conclusions}
\label{section-conclusions}

We have studied the relation between different forms of
divisibility of dynamical maps and collision models that
stroboscopically simulate such dynamical maps. Our findings are
illustrated by Pauli dynamical maps, which allow a particularly
visual pictorial representation of process trajectories in the
parameter space.

A concept of ultimate CP divisible maps has been introduced:
ultimate CP divisible processes can be understood as ultimate
dynamical maps still simulable by collision models with
\textit{factorized} environment. Ultimate CP divisible semigroups
of Pauli maps are fully characterized, with the interaction
Hamiltonian being specified.

Within the framework of collision models, we have demonstrated
additivity and multiplicativity of time-dependent generators of CP
divisible processes. The environment remains factorized in this
case. Roughly speaking, to realize a weighted sum of generators of
CP divisible maps one has to shuffle individual environments using
tensor product.

Using \textit{correlated} environment states, we have explicitly
constructed a collision model realizing the mixture of CP
divisible maps. The latter technique was used to simulate a new
two-parameter family of eternal CP indivisible maps. This family
represents a mixture of two pure dephasing processes and a skewed
version of the generalized amplitude damping process. Continuing
the rough analogy, a mixture of dynamical maps corresponds to
uniting individual environments via direct sum operation.

Also, we have reviewed general properties of P divisible dynamical
maps. In particular, using a quantum analogue of Sanov's theorem
we have noticed that the probability of confusing two states
monotonically increases in P divisible processes. As far as P
indivisible processes are concerned, we have explicitly
constructed collision models simulating arbitrary Pauli dynamical
maps.

\begin{acknowledgements}
This research was initiated during the first Quantum Physics
Unconference organized in Lapland in 2015 by S.M., M.Z., and Teiko
Heinosaari. The study is supported by Russian Science Foundation
under project No. 16-11-00084 and performed in Moscow Institute of
Physics and Technology. S.M. and J.P. acknowledge financial
support from the Horizon 2020 EU collaborative projects QuProCS
(Grant Agreement No. 641277), the Academy of Finland (Project no.
287750), and the Magnus Ehrnrooth Foundation. M.Z. acknowledges
support of projects QETWORK APVV-14-0878, MAXAP VEGA 2/0173/17 and
GA\v CR No. GA16-22211S.
\end{acknowledgements}


\begin{thebibliography}{99}
\bibitem{holevo-giovannetti-2012}
A. S. Holevo and V. Giovannetti, Quantum channels and their
entropic characteristics, Rep. Prog. Phys. {\bf 75}, 046001
(2012).

\bibitem{palma-1996}
G. M. Palma, K.-A. Suominen, and A. K. Ekert, Quantum computers
and dissipation, Proc. R. Soc. Lond. A {\bf 452}, 567 (1996).

\bibitem{aolita-2015}
L. Aolita, F. de Melo, and L. Davidovich, Open-system dynamics of
entanglement: a key issues review, Rep. Prog. Phys. {\bf 78},
042001 (2015).

\bibitem{rivas-2014}
\'{A}. Rivas, S. F. Huelga, and M. B. Plenio, Quantum
non-Markovianity: characterization, quantification and detection,
Rep. Prog. Phys. {\bf 77}, 094001 (2014).

\bibitem{breuer-2016}
H.-P. Breuer, E.-M. Laine, J. Piilo, and B. Vacchini, Colloquium:
Non-Markovian dynamics in open quantum systems, Rev. Mod. Phys.
{\bf 88}, 021002 (2016).

\bibitem{de-vega-2017}
I. de Vega and D. Alonso, Dynamics of non-Markovian open quantum
systems, Rev. Mod. Phys. {\bf 89}, 015001 (2017).

\bibitem{liu-2011}
B.-H. Liu, L. Li, Y.-F. Huang, C.-F. Li, G.-C. Guo, E.-M. Laine,
H.-P. Breuer, and J. Piilo, Experimental control of the transition
from Markovian to non-Markovian dynamics of open quantum systems,
Nature Phys. {\bf 7}, 931 (2011).

\bibitem{tang-2012}
J.-S. Tang, C.-F. Li, Y.-L. Li, X.-B. Zou, G.-C. Guo, H.-P.
Breuer, E.-M. Laine, and J. Piilo, Measuring non-Markovianity of
processes with controllable system-environment interaction, EPL
(Europhysics Letters) {\bf 97}, 10002 (2012).

\bibitem{chiuri-2012}
A. Chiuri, C. Greganti, L. Mazzola, M. Paternostro, and P
Mataloni, Linear Optics Simulation of Quantum Non-Markovian
Dynamics, Sci. Rep. {\bf 2}, 968 (2012).

\bibitem{groblacher-2015}
S. Gr\"{o}blacher, A. Trubarov, N. Prigge, G. D. Cole, M.
Aspelmeyer, and J. Eisert, Observation of non-Markovian
micromechanical Brownian motion, Nature Comm. {\bf 6}, 7606
(2015).

\bibitem{bernardes-2015}
N. K. Bernardes, A. Cuevas, A. Orieux, C.H. Monken, P. Mataloni,
F. Sciarrino, and M. F. Santos, Experimental observation of weak
non-Markovianity, Sci. Rep. {\bf 5}, 17520 (2015).

\bibitem{bernardes-2016}
N. K. Bernardes, J. P. S. Peterson, R. S. Sarthour, A. M. Souza,
C. H. Monken, I. Roditi, I. S. Oliveira, and M. F. Santos, High
Resolution non-Markovianity in NMR, Sci. Rep. {\bf 6}, 33945
(2016).

\bibitem{cialdi-2017}
S. Cialdi, M. A. C. Rossi, C. Benedetti, B. Vacchini, D.
Tamascelli, S. Olivares, and M. G. A. Paris, All-optical quantum
simulator of qubit noisy channels, Appl. Phys. Lett. {\bf 110},
081107 (2017).

\bibitem{breuer-2009}
H.-P. Breuer, E.-M. Laine, and J. Piilo, Measure for the Degree of
Non-Markovian Behavior of Quantum Processes in Open Systems, Phys.
Rev. Lett. {\bf 103}, 210401 (2009).

\bibitem{laine-2010}
E.-M. Laine, J. Piilo, and H.-P. Breuer, Measure for the
non-Markovianity of quantum processes, Phys. Rev. A {\bf 81},
062115 (2010).

\bibitem{wolf-prl-2008}
M. M. Wolf, J. Eisert, T. S. Cubitt, and J. I. Cirac, Assessing
Non-Markovian Quantum Dynamics, Phys. Rev. Lett. {\bf 101}, 150402
(2008).

\bibitem{rivas-2010}
\'{A}. Rivas, S. F. Huelga, and M. B. Plenio, Entanglement and
Non-Markovianity of Quantum Evolutions, Phys. Rev. Lett. {\bf
105}, 050403 (2010).

\bibitem{lorenzo-2013}
S. Lorenzo, F. Plastina, and M. Paternostro, Geometrical
characterization of non-Markovianity, Phys. Rev. A {\bf 88},
020102(R) (2013).

\bibitem{lu-2010}
X.-M. Lu, X. Wang, and C. P. Sun, Quantum Fisher information flow
and non-Markovian processes of open systems, Phys. Rev. A {\bf
82}, 042103 (2010).

\bibitem{luo-2012}
S. Luo, S. Fu, and H. Song, Quantifying non-Markovianity via
correlations, Phys. Rev. A {\bf 86}, 044101 (2012).

\bibitem{bylicka-2014}
B. Bylicka, D. Chru\'{s}ci\'{n}ski, and S. Maniscalco,
Non-Markovianity and reservoir memory of quantum channels: a
quantum information theory perspective, Sci. Rep. {\bf 4}, 5720
(2014).

\bibitem{dhar-2015}
H. S. Dhar, M. N. Bera, and G. Adesso, Characterizing
non-Markovianity via quantum interferometric power, Phys. Rev. A
{\bf 91}, 032115 (2015).

\bibitem{wolf-2008}
M. M. Wolf and J. I. Cirac, Dividing Quantum Channels, Commun.
Math. Phys. {\bf 279}, 147 (2008).

\bibitem{megier-2016}
  N. Megier, D. Chru\'{s}ci\'{n}ski, J. Piilo, and W. T. Strunz,
  Eternal non-Markovianity: from random unitary to Markov chain realisations,
  Sci. Rep. {\bf 7}, 6379 (2017), arXiv:1608.07125.

\bibitem{hall-2014}
M. J. W. Hall, J. D. Cresser, L. Li, and E. Andersson, Canonical
form of master equations and characterization of non-Markovianity,
Phys. Rev. A {\bf 89}, 042120 (2014).

\bibitem{bylicka-2016}
B. Bylicka, M. Johansson, and A. Acin, Constructive Method for
Detecting the Information Backflow of Non-Markovian Dynamics,
Phys. Rev. Lett. {\bf 118}, 120501 (2017), arXiv:1603.04288.

\bibitem{chruscinski-maniscalco-2014}
D. Chru\'{s}ci\'{n}ski and S. Maniscalco, Degree of
Non-Markovianity of Quantum Evolution, Phys. Rev. Lett. {\bf 112},
120404 (2014).

\bibitem{nalezyty-2015}
F. A. Wudarski, P. Nale\.{z}yty, G. Sarbicki, and D.
Chru\'{s}ci\'{n}ski, Admissible memory kernels for random unitary
qubit evolution, Phys. Rev. A {\bf 91}, 042105 (2015).

\bibitem{wudarski-2015}
D. Chru\'{s}ci\'{n}ski and F. A. Wudarski, Non-Markovianity degree
for random unitary evolution, Phys. Rev. A {\bf 91}, 012104
(2015).

\bibitem{chruscinski-siudzinska-2016}
D. Chru\'{s}ci\'{n}ski and K. Siudzi\'{n}ska, Generalized Pauli
channels and a class of non-Markovian quantum evolution, Phys.
Rev. A {\bf 94}, 022118 (2016).

\bibitem{hall-2008}
M. J. W. Hall, Complete positivity for time-dependent qubit master
equations, J. Phys. A: Math. Theor. {\bf 41}, 205302 (2008).

\bibitem{breuer-petruccione-2002}
H.-P. Breuer and F. Petruccione, {\it The Theory of Open Quantum
Systems} (Oxford University Press, Oxford, 2002).

\bibitem{rau-1963}
J. Rau, Relaxation phenomena in spin and harmonic oscillator
systems, Phys. Rev. {\bf 129}, 1880 (1963).

\bibitem{scarani-2002}
V. Scarani, M. Ziman, P. \v{S}telmachovi\v{c}, N. Gisin, and V.
Bu\v{z}ek, Thermalizing Quantum Machines: Dissipation and
Entanglement, Phys. Rev. Lett. {\bf 88}, 097905 (2002).

\bibitem{ziman-2002}
M. Ziman, P. \v{S}telmachovi\v{c}, V. Bu\v{z}ek, M. Hillery, V.
Scarani, and N. Gisin, Diluting quantum information: An analysis
of information transfer in system-reservoir interactions, Phys.
Rev. A {\bf 65}, 042105 (2002).

\bibitem{ziman-buzek-2011}
M. Ziman and V. Bu\v{z}ek, in {\it Quantum Dynamics and
Information}, edited by R. Olkiewicz {\it et al}. (World
Scientific, Singapore, 2011), pp. 199-227.

\bibitem{kretschmann-2005}
D. Kretschmann and R.F. Werner, Quantum channels with memory,
Phys. Rev. A {\bf 72}, 062323 (2005).

\bibitem{giovannetti-2005}
V. Giovannetti, A dynamical model for quantum memory channels, J.
Phys. A: Math. Gen. {\bf 38}, 10989 (2005).

\bibitem{giovannetti-2012}
V. Giovannetti and G. M. Palma, Master Equations for Correlated
Quantum Channels, Phys. Rev. Lett. {\bf 108}, 040401 (2012).

\bibitem{caruso-2014}
F. Caruso, V. Giovannetti, C. Lupo, and S. Mancini, Quantum
channels and memory effects, Rev. Mod. Phys. {\bf 86}, 1203
(2014).

\bibitem{rybar-2012}
T. Ryb\'{a}r, S. N. Filippov, M. Ziman, and V. Bu\v{z}ek,
Simulation of indivisible qubit channels in collision models, J.
Phys. B: At. Mol. Opt. Phys. {\bf 45}, 154006 (2012).

\bibitem{bernardes-2014}
N. K. Bernardes, A. R. R. Carvalho, C. H. Monken, and M. F.
Santos, Environmental correlations and Markovian to non-Markovian
transitions in collisional models, Phys. Rev. A {\bf 90}, 032111
(2014).

\bibitem{bernardes-2017}
N. K. Bernardes, A. R. R. Carvalho, C. H. Monken, and M. F.
Santos, Coarse graining a non-Markovian collisional model, Phys.
Rev. A {\bf 95}, 032117 (2017).

\bibitem{dabrowska-2017}
A. D\k{a}browska, G. Sarbicki, and D. Chru\'{s}ci\'{n}ski, Quantum
trajectories for a system interacting with environment in a single
photon state: counting and diffusive processes, arXiv: 1706.07967.

\bibitem{pellegrini-2009}
C. Pellegrini and F. Petruccione, Non-Markovian quantum repeated
interactions and measurements, J. Phys. A: Math. Theor. {\bf 42},
425304 (2009).

\bibitem{bodor-2013}
A. Bodor, L. Di\'{o}si, Z. Kallus, and T. Konrad, Structural
features of non-Markovian open quantum systems using quantum
chains, Phys. Rev. A {\bf 87}, 052113 (2013).

\bibitem{ciccarello-2013}
F. Ciccarello, G. M. Palma, and V. Giovannetti,
Collision-model-based approach to non-Markovian quantum dynamics,
Phys. Rev. A {\bf 87}, 040103(R) (2013).

\bibitem{ciccarello-ps-2013}
F. Ciccarello and V. Giovannetti, A quantum non-Markovian
collision model: incoherent swap case, Phys. Scr. {\bf T 153},
014010 (2013).

\bibitem{mccloskey-2014}
R. McCloskey and M. Paternostro, Non-Markovianity and
system-environment correlations in a microscopic collision model,
Phys. Rev. A {\bf 89}, 052120 (2014).

\bibitem{kretschmer-2016}
S. Kretschmer, K. Luoma, and W. T. Strunz, Collision model for
non-Markovian quantum dynamics, Phys. Rev. A {\bf 94}, 012106
(2016).

\bibitem{budini-2013}
A. A. Budini, Embedding non-Markovian quantum collisional models
into bipartite Markovian dynamics, Phys. Rev. A {\bf 88}, 032115
(2013).

\bibitem{lorenzo-2016}
S. Lorenzo, F. Ciccarello, and G. M. Palma, Class of exact
memory-kernel master equations, Phys. Rev. A {\bf 93}, 052111
(2016).

\bibitem{hartmann-2005}
L. Hartmann, J. Calsamiglia, W. D\"{u}r, and H. J. Briegel, Spin
gases as microscopic models for non-Markovian decoherence, Phys.
Rev. A {\bf 72}, 052107 (2005).

\bibitem{koniorczyk-2008}
M. Koniorczyk, \'{A}. Varga, P. Rap\v{c}an, and V. Bu\v{z}ek,
Quantum homogenization and state randomization in semiquantal spin
systems, Phys. Rev. A {\bf 77}, 052106 (2008).

\bibitem{vacchini-2016}
B. Vacchini, Generalized Master Equations Leading to Completely
Positive Dynamics, Phys. Rev. Lett. {\bf 117}, 230401 (2016).

\bibitem{diosi-2012}
L. Di\'{o}si, Non-Markovian open quantum systems: Input-output
fields, memory, and monitoring, Phys. Rev. A {\bf 85}, 034101
(2012).

\bibitem{lorenzo-2017}
S. Lorenzo, F. Ciccarello, and G. M. Palma, Composite quantum
collision models, arXiv:1705.03215.

\bibitem{cusumano-2017}
S. Cusumano, A. Mari, and V. Giovannetti, Interferometric quantum
cascade systems, Phys. Rev. A {\bf 95}, 053838 (2017).

\bibitem{lorenzo-2-2017}
S. Lorenzo, F. Ciccarello, G. M. Palma, and B. Vacchini, Quantum
non-Markovian piecewise dynamics from collision models,
arXiv:1706.09025.

\bibitem{lupo-2010}
C. Lupo, V. Giovannetti, and S. Mancini, Memory effects in
attenuation and amplification quantum processes, Phys. Rev. A {\bf
82}, 032312 (2010).

\bibitem{jin-2015}
J. Jin, V. Giovannetti, R. Fazio, F. Sciarrino, P. Mataloni, A.
Crespi, and R. Osellame, All-optical non-Markovian stroboscopic
quantum simulator, Phys. Rev. A {\bf 91}, 012122 (2015).

\bibitem{lorenzo-2015}
S. Lorenzo, R. McCloskey, F. Ciccarello, M. Paternostro, and G. M.
Palma, Landauer's Principle in Multipartite Open Quantum System
Dynamics, Phys. Rev. Lett. {\bf 115}, 120403 (2015).

\bibitem{attal-2006}
S. Attal and Y. Pautrat, From Repeated to Continuous Quantum
Interactions, Ann. Henri Poincar\'{e} {\bf 7}, 59 (2006).

\bibitem{kolodynski-2017}
J. Ko{\l}ody\'{n}ski, J. Bohr Brask, M. Perarnau-Llobet, and B.
Bylicka, Adding dynamical generators in quantum master equations,
arXiv:1704.08702.

\bibitem{benatti-2017}
F. Benatti, D. Chru\'{s}ci\'{n}ski, and S. Filippov, Tensor power
of dynamical maps and positive versus completely positive
divisibility, Phys. Rev. A {\bf 95}, 012112 (2017).

\bibitem{ruskai-2002}
M. B. Ruskai, S. Szarek, and E. Werner, An analysis of
completely-positive trace-preserving maps on $M_2$, Linear Algebra
Appl. {\bf 347}, 159 (2002).

\bibitem{bengtsson-2006}
I. Bengtsson and K. \.{Z}yczkowski, {\it Geometry of Quantum
States. An Introduction to Quantum Entanglement} (Cambridge
University Press, New York, 2006).

\bibitem{gks-1976}
V. Gorini, A. Kossakowski, and E. C. G. Sudarshan, Completely
positive dynamical semigroups of $N$-level systems, J. Math. Phys.
{\bf 17}, 821 (1976).

\bibitem{lindblad-1976}
G. Lindblad, On the Generators of Quantum Dynamical Semigroups,
Commun. Math. Phys. {\bf 48}, 119 (1976).

\bibitem{nielsen-2000}
M. A. Nielsen and I. L. Chuang, {\it Quantum Computation and
Quantum Information} (Cambridge University Press, Cambridge,
2000).

\bibitem{luchnikov-2017}
I. A. Luchnikov and S. N. Filippov, Quantum evolution in the
stroboscopic limit of repeated measurements, Phys. Rev. A {\bf
95}, 022113 (2017).

\bibitem{werner-2008}
D. Kretschmann, D. Schlingemann, R. F. Werner, A continuity
theorem for Stinespring's dilation, Journal of Functional Analysis
{\bf 255}, 1889 (2008).

\bibitem{li-2017}
L. Li and M. J. W. Hall, Eternal non-Markovianity is generic for
the spin-boson model, arXiv:1701.01292.

\bibitem{muller-hermes-2017}
A. M\"{u}ller-Hermes and D. Reeb, Monotonicity of the Quantum
Relative Entropy Under Positive Maps, Ann. Henri Poincare {\bf
18}, 1777 (2017).

\bibitem{vedral-1997}
V. Vedral, M. B. Plenio, K. Jacobs, and P. L. Knight, Statistical
inference, distinguishability of quantum states, and quantum
entanglement, Phys. Rev. A {\bf 56}, 4452 (1997).

\bibitem{holevo-1998}
A. S. Holevo, Quantum coding theorems, Russ. Math. Surveys {\bf
53}, 1295 (1998).

\bibitem{horodecki-2003}
M. Horodecki, P. W. Shor, and M. B. Ruskai, Entanglement Breaking
Channels, Rev. Math. Phys. {\bf 15}, 629 (2003).

\bibitem{moravcikova-ziman-2010}
L. Morav\v{c}\'{i}kov\'{a} and M. Ziman, Entanglement-annihilating
and entanglement-breaking channels, J. Phys. A: Math. Theor. {\bf
43}, 275306 (2010).

\bibitem{filippov-rybar-ziman-2012}
S. N. Filippov, T. Ryb\'{a}r, and M. Ziman, Local two-qubit
entanglement-annihilating channels, Phys. Rev. A {\bf 85}, 012303
(2012).

\bibitem{muller-hermes-2016}
A. M\"{u}ller-Hermes, D. Reeb, and M. M. Wolf, Positivity of
linear maps under tensor powers, J. Math. Phys. {\bf 57}, 015202
(2016).

\bibitem{filippov-magadov-2017}
S. N. Filippov and K. Yu. Magadov, Positive tensor products of
maps and $n$-tensor-stable positive qubit maps, J. Phys. A: Math.
Theor. {\bf 50}, 055301 (2017).

\end{thebibliography}
\end{document}